\documentclass[twocolumn,aps,prl,showpacs,superscriptaddress,final]{revtex4-1}             
\usepackage{bm}
\usepackage{hyperref}
\hypersetup{
	colorlinks = true,
	linkcolor  = blue,
	citecolor  = blue,
	filecolor  = magenta,
	urlcolor   = cyan
}
\usepackage [english]{babel}
\usepackage [utf8]{inputenc}
\usepackage [T1]{fontenc}
\usepackage{graphicx}
\usepackage{amsmath}
\usepackage{amssymb}
\usepackage{siunitx}
\usepackage{booktabs}
\usepackage{array}
\usepackage[normalem]{ulem}
\usepackage[dvipsnames]{xcolor}
\definecolor{royalblue}{HTML}{4169e1}

\definecolor{bond}{HTML}{007007}

\providecommand{\ifhighlighting}{\iffalse} %
\ifhighlighting
\usepackage{xcolor}
\newcommand{\hlstart}{\color{purple}}
\newcommand{\hlend}{\color{black}}
\else
\newcommand{\hlstart}{}
\newcommand{\hlend}{}
\fi

\newcommand{\be}{\begin{equation}}
	\newcommand{\e}{\end{equation}}

\newcommand{\beml}{\begin{subequations}}
	\newcommand{\eml}{\end{subequations}}
\newcommand{\beq}{\begin{eqnarray}}
	\newcommand{\eq}{\end{eqnarray}}
\newcommand{\ba}{\begin{array}}
	\newcommand{\ea}{\end{array}}

\usepackage{braket}

\begin{document}
	\date{February 16, 2023}
	
	\title{Highly-transmitting modes of light in dynamic atmospheric turbulence}
	
	\author{David Bachmann}
	
	\affiliation{Physikalisches Institut, Albert-Ludwigs-Universit\"at Freiburg, Hermann-Herder-Str.\ 3,
		D-79104 Freiburg, Germany}
	
	\author{Mathieu Isoard}
	
	\affiliation{Physikalisches Institut, Albert-Ludwigs-Universit\"at Freiburg, Hermann-Herder-Str.\ 3,
		D-79104 Freiburg, Germany}
	
	\affiliation{Current address: Laboratoire Kastler Brossel, Sorbonne Université, ENS-Université PSL, 
		Collège de France, CNRS; 4 place Jussieu, F-75252 Paris, France}
	
	\author{Vyacheslav Shatokhin}
	
	\affiliation{Physikalisches Institut, Albert-Ludwigs-Universit\"at Freiburg, Hermann-Herder-Str.\ 3,
		D-79104 Freiburg, Germany}
	
	\affiliation{EUCOR Centre for Quantum Science and Quantum Computing, Albert-Ludwigs-Universit\"at Freiburg, Hermann-Herder-Str.3, D-79104 Freiburg, Germany}
	
	\author{Giacomo Sorelli}

	\affiliation{Laboratoire Kastler Brossel, Sorbonne Université, ENS-Université PSL, 
		Collège de France, CNRS; 4 place Jussieu, F-75252 Paris, France}
	
	\affiliation{Current address: Fraunhofer Institute for Optronics, System Technology and Image Exploitation - IOSB,
		Gutleuthausstrasse 1, 76275 Ettlingen, Germany}

	\author{Nicolas Treps}
	
	\affiliation{Laboratoire Kastler Brossel, Sorbonne Université, ENS-Université PSL, 
		Collège de France, CNRS; 4 place Jussieu, F-75252 Paris, France}
	
	\author{Andreas Buchleitner}
	
	\affiliation{Physikalisches Institut, Albert-Ludwigs-Universit\"at Freiburg, Hermann-Herder-Str.\ 3,
		D-79104 Freiburg, Germany}
	
	\affiliation{EUCOR Centre for Quantum Science and Quantum Computing, Albert-Ludwigs-Universit\"at Freiburg, Hermann-Herder-Str.3, D-79104 Freiburg, Germany}
	
	\begin{abstract}
		\noindent
		\hlstart
		
		We show that instantaneous spatial singular modes of light in a dynamically evolving, turbulent atmosphere offer significantly improved high-fidelity signal transmission as compared to standard encoding bases corrected by adaptive optics.
		Their enhanced stability in stronger turbulence is associated with a subdiffusive algebraic decay of the transmitted power with evolution time.
		
		\hlend
		\noindent
		\newline
		DOI: \href{https://doi.org/10.1103/PhysRevLett.130.073801}{10.1103/PhysRevLett.130.073801}
	\end{abstract}
	
	\maketitle
	
	\emph{Introduction.---}
	\label{sec:intro}
	\hlstart
	Wave transport in random scattering media is ubiquitous in communication, sensing and imaging, from astronomical \cite{Roddier04} over mesoscopic \cite{Labeyrie99,Popoff14,Hsu17} to microscopic scales \cite{DOROKHOV1984}.
	In all these rather diverse physical settings, the common goal is to faithfully transmit and filter relevant information generated by the sender, probed by the scattering wave, or emitted by some unknown object(s). 
	To achieve an efficient transmission and retrieval of information, it is indispensable to mitigate the random, i.e., uncontrolled modulation of the transmitted signal's phase and intensity profile.
	
	In the atmosphere, the primary sources of uncontrolled light modulations are turbulent eddies, i.e. blobs of air with smoothly varying refractive index. On the one hand, the minimal size of turbulent eddies is much larger than optical wavelengths \cite{Roggemann97}, accordingly scattering on such eddies occurs mainly in the forward direction \cite{Fante75}. On the other hand, the eddies' sizes are comparable to the typical transverse width of propagating beams \cite{Roggemann97}, resulting in random phase shifts of the light's transverse profile \cite{Fante75}. Furthermore, a combination of refraction and diffraction gives rise to intensity fluctuations upon transmission \cite{Fante75}. Although phase errors can be compensated by adaptive optics (AO) \cite{Roggemann97, Roddier04, Ren14, Ren16, Zhao20, Tyson16}, the latter cannot alleviate intensity fluctuations \cite{Sorelli19}. Nor can AO reduce transmission losses arising due to finite-size apertures.

	We propose
	to exploit the medium's intrinsic properties for signal transmission across random disorder:
	Wave propagation in static disordered media generically allows for the formation of highly-transmitting spatial modes \cite{Rotter17,Segev13,Yi19}.
	These modes qualify due to their minimal transverse losses as ideal candidates for high-dimensional signal encoding.
	However, their stability
	is challenged in a \emph{dynamically evolving} medium -- here the Earth's atmosphere.
	Yet, even in the time-dependent atmosphere the channel's geometry and turbulence parameters 
	remain invariant. Therefore, one can expect that certain robust features are inscribed into instantaneous modes lending them persistent stability properties.
	Our present purpose is to consolidate this expectation, and to quantify the time scales over which such highly-transmitting modes offer resilient signal transmission with reduced losses.

\begin{figure}
		\includegraphics[width=\columnwidth]{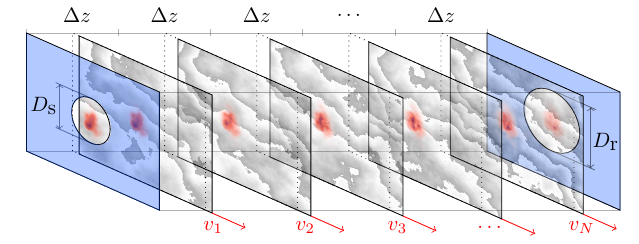}
		\caption{
			Horizontal transmission channel of fixed length $L$ across a dynamically evolving, turbulent atmosphere:
			The medium (confined between coaxial circular source and receiver apertures of diameters $D_\text{s}$ and $D_\text{r}$, respectively) is represented by a collection of equidistant phase screens.
			The turbulence strength governs their transverse structure and their number  $L/\Delta z$ is determined by requiring weak intensity fluctuations at every propagation step. Temporal evolution is modeled by individual shifting of phase screens according to wind velocities $v_i$ (red arrows).
		}
		\label{fig:model}
	\end{figure}

	\emph{The model.---}
	\label{sec:model}
	We consider the horizontal propagation of a monochromatic laser beam through  a clear  atmospheric channel (e.g., free of fog or clouds), limited by two coaxial circular source and receiver apertures with diameters $D_\text{s}$ and $D_\text{r}$, respectively, see Fig.~\ref{fig:model}.
	We study
	the spatial field distribution of the propagating beam, and choose a wavelength of $\lambda=1550\,$nm which is in the infrared transparency window of the Earth's atmosphere \cite{Kopeika98,Andrews05}.
	Furthermore, we consider propagation distances much shorter than the light's transport mean free path \cite{Carminati21,meanfreepath}, thereby neglecting the small attenuation of the field due to molecular and aerosol absorption and scattering \cite{Andrews05,Kopeika98}.
	In this case, the propagation of a scalar light wave $\psi({\bf r}=x,y,z)$ obeys the \emph{stochastic parabolic equation} \cite{Andrews05}
	\begin{equation}
		- 2ik\,\frac{\partial \psi(\bf{r})}{\partial z} =  \Delta_\perp \psi({\bf r}) + 2k^2\,\delta n({\bf r}, t)\, \psi({\bf r}),
		\label{eq:spe}
	\end{equation}
	where $k=2\pi/\lambda$, $\Delta_\perp$ -- the transverse Laplace operator with respect to the propagation axis $z$ -- accounts for wave diffraction, and $\delta n({\bf r},t)$ is the fluctuating part of the refractive index of air, to model the random turbulent potential and its time dependence.
	
	The statistical properties of $\delta n({\bf r},t)$ in Eq.~\eqref{eq:spe} are governed, according to \emph{Kolmogorov theory} \cite{Kolmogorov41a}, by the refractive index power spectrum $\Phi_n(\kappa) \sim \kappa^{-11/3}$, where $\kappa$ denotes the transverse spatial frequency of refractive index fluctuations.
	These random inhomogeneities of the refractive index induce phase distortions 
	with typical correlation radius given by the \emph{Fried parameter} $r_0$ \cite{Andrews05}, which is inversely proportional to the turbulence strength. The Fried parameter typically varies on the timescale of hours \cite{Andrews05} and can thus be assumed time-independent.
	Furthermore, phase distortions combined with diffraction give rise to intensity fluctuations, whose strength is quantified by the \emph{Rytov variance} $\sigma_\text{R}^2(r_0, L)$ \cite{Tyson16, Tatarskii16, Ishimaru78}.
	
	For fixed times $t$, the stochastic parabolic Eq.~\eqref{eq:spe} is accurately solved by a \emph{split-step} method \cite{Schmidt10,Lukin02,Sorelli19}, which relies on segmenting the propagation path into discrete, medium-induced phase modulations, i.e., phase screens, equally interconnected by vacuum diffraction (Fig.~\ref{fig:model}).
	Although every elementary propagation step of length $\Delta z$ only introduces phase errors \cite{supp}, 
	the assembled path reliably models as well intensity fluctuations which originate
	from constructive or destructive interference.
	To implement this
	numerically,
	we employ Fourier optics methods \cite{Born99,Goodman05,Schmidt10,Johansson94} and standard phase screens \cite{Schmidt10,Johansson94} augmented by Zernike polynomials \cite{Roddier90}.
	
	Mathematically, the combination of phase distortions and free diffraction can be represented by the unitary operator $U_{\rm turb}$ acting on the propagating wave; the geometric truncation due to source and receiver apertures (Fig.~\ref{fig:model}) are captured by projection operators $\Pi_\text{s}$ and $\Pi_\text{r}$, respectively.
	Combined, this defines the channel's \emph{turbulence operator} $T_{\rm turb}=\Pi_\text{r}\,U_{\rm turb}\,\Pi_\text{s}$ which maps modes from the source space $\mathcal{H}_\text{s}$ onto the receiver space $\mathcal{H}_\text{r}$.
	In general, this operator is non-unitary, since light escapes the receiver aperture due to diffraction, turbulence-induced broadening and beam wandering \cite{Andrews05}.

	By performing a singular value decomposition (SVD) \cite{Rotter17,Miller19}  of $T_{\rm turb}$, we find its orthonormal transmission channels:
	\emph{Source} modes $v_{s}(\boldsymbol{\rho})\in\mathcal{H}_\text{s}$ with mode index $s$ are coupled bijectively to \emph{receiver} modes $u_{s}(\boldsymbol{\rho}^\prime)\in\mathcal{H}_\text{r}$, each associated with \emph{singular values} $\tau_s$ quantifying the transmitted power per channel, where $\boldsymbol{\rho}\in D_\text{s}$ and $\boldsymbol{\rho}^\prime\in D_\text{r}$ are transverse position vectors in the source and receiver apertures.
	The action of $T_{\rm turb}$ on an input state $\Psi\in\mathcal{H}_\text{s}$ thus has the explicit form 
	\begin{equation}
		(T_{\rm turb}\star \, \Psi)(\boldsymbol{\rho}') = \sum_{s=0}^{S-1} \sqrt{\tau_s} \, u_{s}(\boldsymbol{\rho}^\prime) \, \langle v_{s}(\boldsymbol{\rho}), \Psi(\boldsymbol{\rho}) \rangle\, ,
		\label{eq:svd}
	\end{equation}
	where $\star$ denotes the convolution, $\langle.\, ,.\rangle$ stands for the standard scalar product in transverse space, and the number $S$ of singular modes with non-vanishing weight depends on the source and receiver apertures' cross sections.
	Practically, the SVD is applied to a matrix representation of the turbulence operator, known as \emph{transmission matrix}
	$(T_{\rm turb})_{ij} = \langle \phi_i(\boldsymbol{\rho}^\prime),\, (T_{\rm turb}\star \, \psi_j)(\boldsymbol{\rho'})\rangle$,
	where $\psi_j(\boldsymbol{\rho})\in\mathcal{H}_\text{s}$ and $\phi_i(\boldsymbol{\rho}^\prime)\in\mathcal{H}_\text{r}$ are basis modes on the source and receiver side, respectively.
	For an accurate representation of transmitted fields confined by circular apertures, we choose \emph{Laguerre-Gaussian} (LG) modes at the source side, which also feature convenient diffraction properties \cite{Andrews05}.
	To resolve the fine details of turbulence-induced distortions imprinted into the receiver modes a large number $Q\gg S$ of pixel modes is employed at the receiver's end. 
	All parameters of our model \cite{ourchannel} are chosen to match realistic experimental conditions \cite{Lavery17, Krenn15}.

\begin{figure}
		\includegraphics[width=\columnwidth]{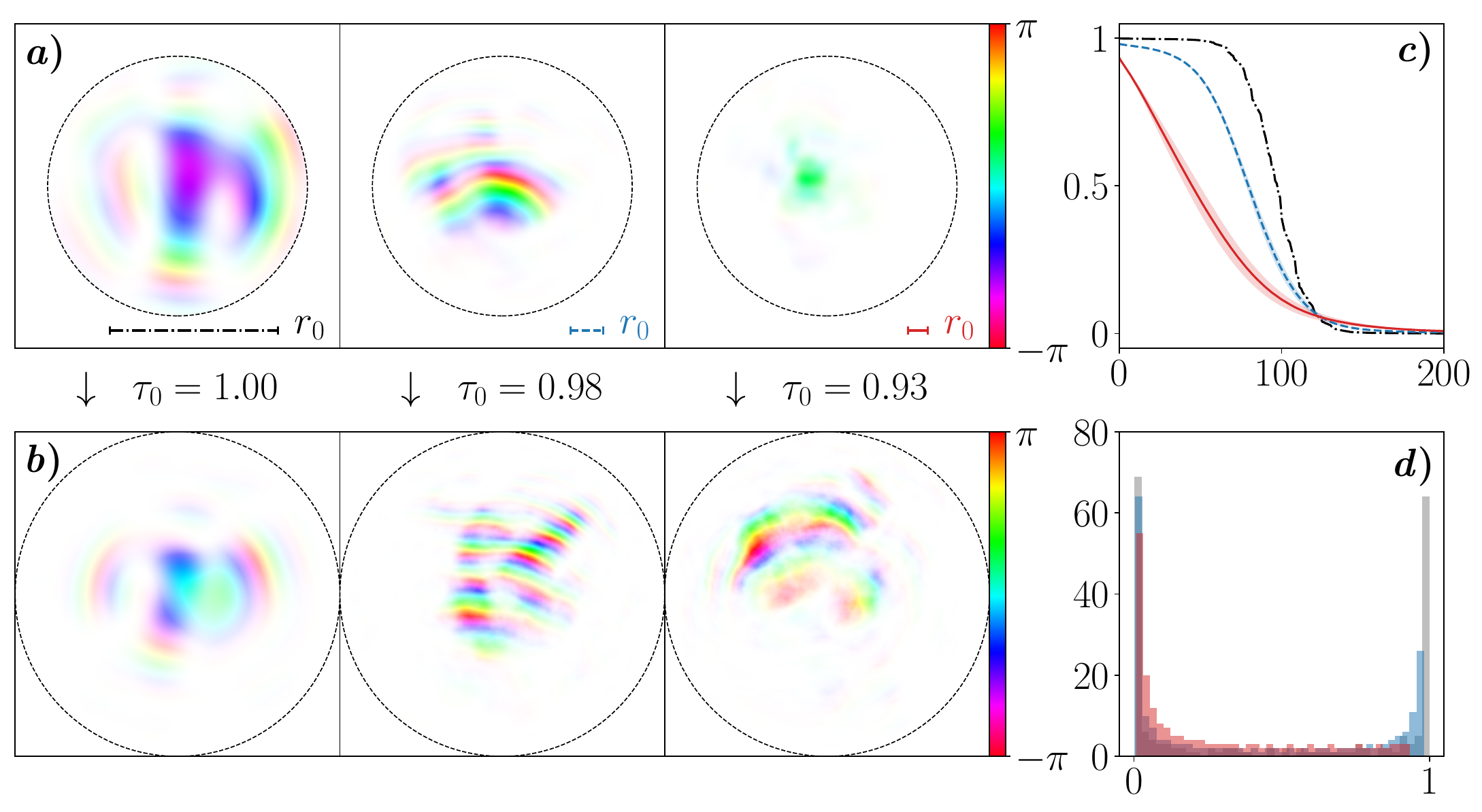}
		\caption{
			Transverse profiles (phase is color-coded and intensity is plotted as opacity)
			of instantaneous source (a) and receiver (b) singular modes associated with the largest singular value $\tau_0$ of a \emph{single} realization of a turbulent medium with Fried parameters $r_0=130, 25, 15\,$mm, corresponding to Rytov variances $\sigma_\text{R}^2=0.18, 4.16, 6.72$ (left to right) in our channel \cite{ourchannel}.
			(c) distributions of the average
			singular values $\overline{\tau}_s$ in decreasing order, each obtained by averaging over 500 disorder realizations (error bands give one standard deviation) with same parameters and color code as in (a,b). (d) corresponding histograms. 
		}
		\label{fig:modes}
	\end{figure}

	\emph{Properties of instantaneous singular modes.---}
	\label{sec:frozen}
	Before considering dynamical turbulence, we investigate the statistical properties of singular modes associated with an \emph{instantaneous} atmospheric realization.
	The average distribution of singular values over an ensemble of random realizations is known in the asymptotic regime of weak turbulence \cite{Shapiro74} and aligns with our results.
	But our present numerical approach captures the properties of the singular modes under general, non-asymptotic turbulence conditions, and thus considerably expands over earlier studies \cite{Shapiro74,Kahn18,Borcea20,Shatokhin20}. 
	Representative examples of intensity and phase distributions of highly-transmitting (i.e., associated singular values $\tau_s$ close to one) source (a) and receiver (b) singular modes are shown in Fig.~\ref{fig:modes}, for three values of $r_0$ corresponding to the regimes of weak, moderate and strong intensity fluctuations (left to right).
	We note that with increasing turbulence strength, the source modes occupy smaller transverse areas.
	In addition, non-trivial phases of both, the source and receiver modes [encoded in the color gradient in Fig. \ref{fig:modes}(a,b)] are another salient feature -- highlighting the relevance of constructive interference effects for effective transmission.
	
	The right-hand panels of Fig.~\ref{fig:modes} show the distribution of singular values (c) and its histogram (d) after \emph{ensemble averaging} over 500 atmospheric realizations.
	The black dash-dot curve corresponds to weak intensity fluctuations initially leveling for approximately 80 perfectly-transmitting singular modes associated with $\overline{\tau}_s= 1$ (the overline designates the ensemble average).
	The performance of further singular modes drops quickly towards vanishing transmittance, i.e., $\overline{\tau}_s= 0$, where the wiggles in the curve 
	reflect residual degeneracy of singular values stemming from geometric symmetries in the absence of turbulence \cite{Miller19}.
	This sharp separation between
	open ($\overline{\tau}_s= 1$) and closed channels ($\overline{\tau}_s= 0$)
	corresponds to a bimodal distribution [Fig.~\ref{fig:modes}(d)] and agrees with the singular value distribution observed in the absence of \cite{Slepian65} or in weak turbulence \cite{Shapiro74}, as well as in other complex scattering systems \cite{DOROKHOV1984,NazarovPRL1994,BennakerRevModPhys}.
	However,
	with increasing turbulence strength
	the open channels evolve into partially transmitting channels, i.e., $\overline{\tau_s}<1$ \cite{GoetschyPRL2013,Popoff14,Hsu17}.
	With increasing turbulence strength the open-channel plateau shrinks (blue dashed line in Fig.~\ref{fig:modes} for moderate intensity fluctuations) and eventually turns into a monotonic decay (red solid line) of the transmittance for strong intensity fluctuations.
	In the latter regime, the resulting unimodal singular value distribution [Fig.~\ref{fig:modes}(d)] is dominated by closed channels. Notwithstanding, even in this case, we identify more than ten modes with singular values $\overline{\tau}_s>0.9$.

	\emph{Stability in dynamic turbulence.---}
	Given the non-negligible set of highly-transmitting modes -- even 
	in strong
	turbulence -- we need to address their stability under the atmosphere's dynamical evolution.
	To account for the latter, we adopt \emph{Taylor's hypothesis} \cite{Taylor35}, which states that
	the short time ($t\lesssim 1\,$s) atmospheric evolution is described by the wind-induced transverse flow of static turbulent eddies \cite{Andrews05}.
	Within the framework of our split-step approach, this behavior is implemented by transverse shifts of individual phase screens with respect to the propagation direction according to appropriate velocity distributions $V$ \cite{Lukin02,Anzuola17} (cf. Fig.~\ref{fig:model}).
	Here, we assume normally-distributed transverse winds with mean $\langle V \rangle$ and variance $\Delta V^2$ which induce atmospheric coherence times
	$t_{\rm c}=6.88^{-3/5}\,r_0/\langle V\rangle$ \cite{Roddier04}.
	
	In general, the atmospheric propagation of highly-transmitting modes $v_s(\boldsymbol{\rho})$ associated with some initial realization of turbulence at $t=0$ will yield some unknown field $f_s(\boldsymbol{\rho}^\prime,t)$ on the receiver's end for $t>0$.
	As seen from Eq.~(\ref{eq:svd}), this field has to satisfy the initial condition 
	$f_s(\boldsymbol{\rho}^\prime,0)\equiv \sqrt{\tau_s}\, u_s(\boldsymbol{\rho}^\prime)$.
	Consequently, the mean \emph{transmitted power} is given by
	$P_s(t) := \overline{| f_s(\boldsymbol{\rho}^\prime,t) |^2}$
	with $P_s(0)=\overline{\tau}_s$,
	and a mode's stability may be characterized by the 
	way
	$P_s(t)$ decays.
	Moreover, the evolution of the mean \emph{crosstalk} matrix, $C_{s',s}(t):=\overline{|\langle f_{s'}(\boldsymbol{\rho}^\prime,t), u_s(\boldsymbol{\rho}^\prime) \rangle|}$ quantifies the dynamically induced, undesired transfer of power from $u_s(\boldsymbol{\rho}^\prime)$ into other output modes.
	We benchmark this stability against the performance of optimized LG modes
	\footnote{The LG beam waist $w_0=29.7\,$mm is chosen to maximize transmission in vacuum. For the definitions of transmitted power $P_s(t)$ and crosstalk $C_{s',s}(t)$ of LG modes $s$ denotes the azimuthal index $\ell=0,\pm 1,\dots,\pm 5$ and
		$f_{s'}(\boldsymbol{\rho}^\prime,t)$ is given by modes propagated through evolved turbulence at time $t$ while $u_s(\boldsymbol{\rho}^\prime)$ corresponds to LG modes propagated through vacuum.}
	subject to ideal AO corrections \cite{Sorelli19}. This means all phase errors attained by a propagated plane wave are subtracted form the received LG modes.

	\begin{figure}
		\includegraphics[width=\columnwidth]{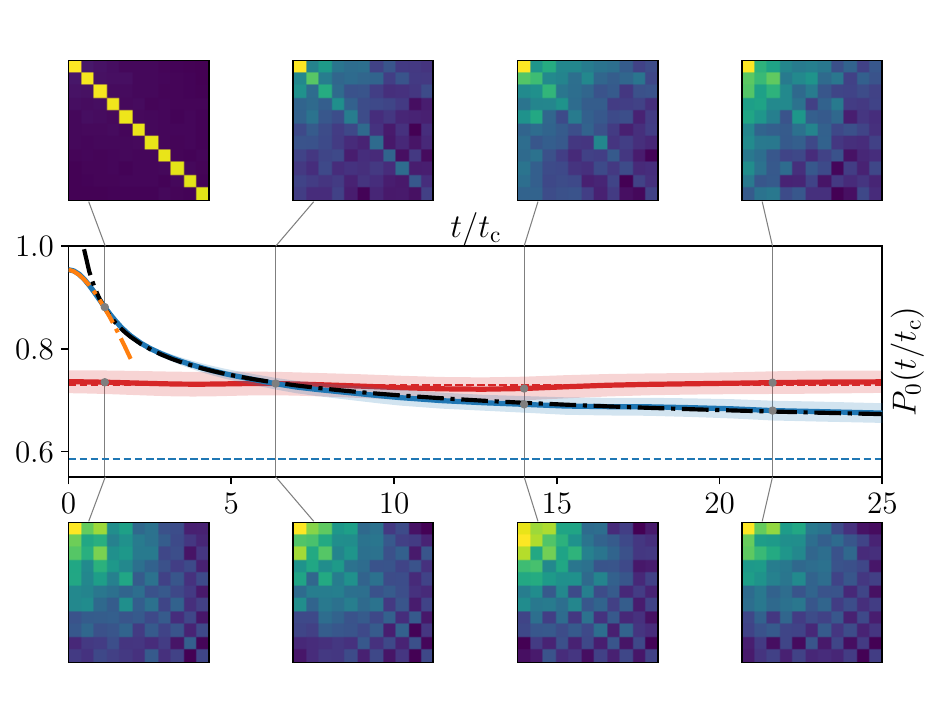}
		\caption{
			Mean transmitted power (main panel) and crosstalk matrices \cite{ourcrosstalk} (top and bottom rows) of highest-transmitting ($s=0$) singular modes (blue, top) and optimized LG modes (red, bottom) versus $t/t_\text{c}$ for strong intensity fluctuations ($\sigma_\text{R}^2=6.72$, $r_0=15$\,mm; same channel geometry as previously \cite{ourchannel}) with $t_\text{c}=1.57\,$ms \cite{ourwind}
			(error bands give one standard deviation of 150 realizations).
			The singular modes' power transmittance was identified as an initial exponential decay (orange) for $t\lesssim t_\text{x}$ saturating algebraically (black) for $t>t_\text{x}$, cf. Eq.~\eqref{eq:Pt} (here $t_\text{x}\approx t_\text{c}$).
			Dashed horizontal lines (same color coding) represent transmission through uncorrelated turbulence coinciding with the independently found asymptotic limit of the algebraic decay for singular modes.}
		\label{fig:power}
	\end{figure}
	
	The main panel of Fig.~\ref{fig:power} compares the mean transmitted power of the singular mode with the highest transmittance ($s=0$; blue) to the AO corrected Gaussian (i.e., LG, $p=\ell=0$) mode (red) in the regime of strong intensity fluctuations, i.e., $\sigma_\text{R}^2=6.72$.
	The transmitted power is plotted as a function of time $t$ in units of $t_\text{c}$, which
	renders the temporal dynamics of power transmittance and crosstalk independent of the mean wind speed $\langle V \rangle$. 
	First, we observe that the Gaussian
	mode (red line in Fig.~\ref{fig:power}), being independent of the turbulent medium, transfers the same power on average.
	In contrast, being optimized for the initial realization of the turbulent medium, singular modes transmit less and less power as the atmosphere evolves.
	Nevertheless, it is evident from Fig.~\ref{fig:power} that the singular mode with the largest singular value $\tau_0$ 
	outperforms the 
	power transmission of the Gaussian mode for sufficiently short time scales $t<t_\text{int} \simeq 6\, t_\text{c}$, where $t_\text{int}$ denotes the time when the transmitted powers of singular modes and LG modes intersect.

	We establish by careful fits that the singular modes' transmitted power is governed by the following decay law:
	\begin{equation}
		P_s(t/t_\text{c})= \left\{
		\begin{array}{ll}
			\overline{\tau}_s \exp\left[- (a_s\,t/t_\text{c})^{5/3}\right] & \quad t \lesssim t_\text{x}\\
			b_s\,(t/t_\text{c})^{-c_s} + \bar{\delta}_s & \quad t > t_\text{x}, \\
			
		\end{array}
		\right.
		\label{eq:Pt}
	\end{equation}
	where the mean transmitted power at $t = 0$ is given by the averaged singular values $\bar{\tau}_s$, while $\bar{\delta}_s$ represents the asymptotic limit $t \to \infty$,
	obtained by propagating the singular modes through uncorrelated (rather than wind-shifted, see Fig.~\ref{fig:model}) phase screens.
	The parameters $a_s$, $b_s$ and $c_s$ (for $s= 0, \dots, 10$) are instead determined through non-linear fits for different turbulence strengths \cite{supp}. 
	In particular, $a_s$ ($b_s$ and $c_s$) are obtained by fitting the numerical data for $t < t_\text{x}$ ($t > t_\text{x}$), where the \emph{crossover time} $t_\text{x}$ is
	the inflection point of $P_s(t/t_\text{c})$, which is inversely proportional to the mean wind speed, determined by the channel geometry and independent of the mode number $s$ \cite{supp}.
	
	The short-time decay in Eq.~\eqref{eq:Pt} 
	is similar to the known decay law 
	$\sim\exp[-(t/t_\text{c})^{5/3}]$ of the \emph{Strehl ratio} \cite{Roddier04}, a quantifier of imaging quality in 
	optics \cite{Tyson16,Born99}.
	Notably, we find this initial decay to be universal: $a_s \simeq 0.2$ is independent of the Fried parameter $r_0$, of the mean transverse wind speed $\langle V \rangle$ -- due to rescaling with $t_\text{c}$ -- and of the mode number $s$.
	In other words,
	the initial rate of a mode's transmittance loss is solely determined by the static channel geometry,
	while
	its duration
	$t_\text{x}$ is
	fixed by the mean transverse wind speed $\langle V \rangle$.
	
	The subsequent algebraic decay of transmitted power for $t>t_\text{x}$ is \emph{subdiffusive}, i.e. decays slower than $\sim1/\sqrt{t}$.
	For the eleven highest-transmitting singular modes and various turbulence strengths we find $0.32 \lesssim c_s \lesssim 0.46 < 1/2$ \cite{supp}.
	This slow decay may be attributed to long-range transverse spatial correlations due to the Kolmogorov turbulence power spectrum $\sim\kappa^{-11/3}$ \cite{Kolmogorov41a, Andrews05}:
	By means of Taylor's hypothesis Kolmogorov's long-ranging spatial correlations are translated into slowly decaying temporal correlations inducing the slow subdiffusive decay of transmittance.
	To corroborate this hypothesis, we consider light propagation through a disordered medium
	with only short-range correlations.
	Phase distortions in such a medium may be described by a Gaussian power spectrum \cite{Martin88, Ishimaru78} for which we observe a purely exponential decay of the singular modes' transmitted power \cite{supp}.

	The temporal stability of singular modes as compared to optimized LG modes is further substantiated by the evolution of their crosstalk matrices (top and bottom rows in Fig.~\ref{fig:power}, respectively).
	The coupling among the eleven highest-transmitting singular modes exhibits a prominent diagonal structure for $t\leq t_{\rm int}$, corresponding to negligible crosstalk.
	Furthermore, this structure leaves fingerprints even for longer times, while the considered LG modes
	feature strong crosstalk at all times.

	\begin{figure}
		\includegraphics[width=\columnwidth]{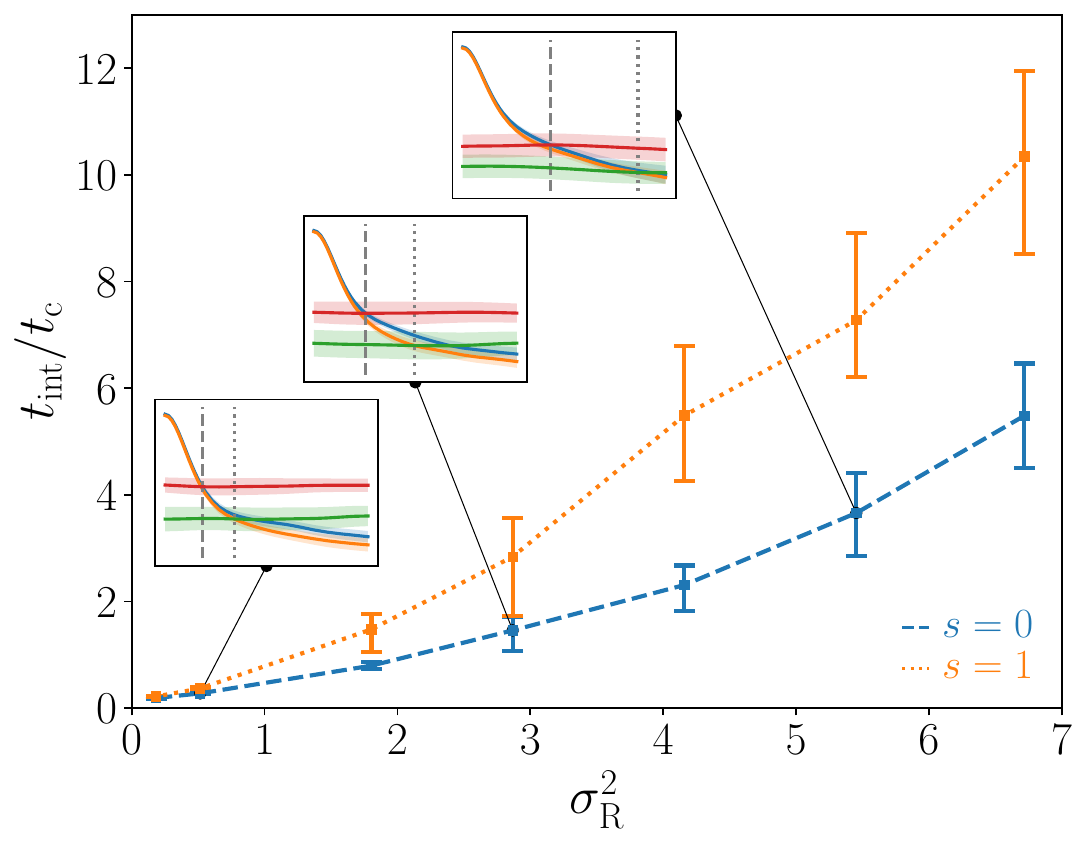}
		\caption{Rescaled average time interval $t_\text{int}/t_\text{c}$ versus Rytov variance during which the highest ($s=0$, blue dashed) and second-highest ($s=1$, orange dotted) transmitting singular modes transfer more power than respective LG modes (error bars give one standard deviation of 150 realizations).
			The insets sketch the average transmitted power of the two highest-transmitting singular ($s=0$ blue, $s=1$ orange) and LG ($s=0$ red, $s=1$ green) modes versus rescaled time (cf. Fig.~\ref{fig:power}) with vertical lines indicating corresponding intersections.}
		\label{fig:gap}
	\end{figure}

	Finally, we assess the effect of intensity fluctuations on the time interval $t_\text{int}$ over which singular modes perform better than LG modes.
	Figure~\ref{fig:gap} shows the extracted intersection time $t_\text{int}/t_\text{c}$ versus Rytov variance $\sigma_\text{R}^2$ for the two highest transmitting modes ($s=0$ in blue and $s=1$ in orange) for the same channel geometry as previously.
	We observe that $t_\text{int}/t_\text{c}$ is a nonlinear, monotonically increasing function of the Rytov variance.
	Since all of the considered eleven singular modes \cite{supp} performed similarly well
	(cf. blue and orange curves for the two highest-transmitting modes in the insets of Fig.~\ref{fig:gap})
	and the transmittance of the LG modes decreases quickly for higher mode numbers (cf. red and green curves in the insets), the growth rate of the intersection time $t_\text{int}/t_\text{c}$ increases with the mode number $s$.
	This is consistent with our above discussion of the distinct crosstalk properties of singular and LG modes (cf. Fig.~\ref{fig:power}).
	In particular, it emphasizes the potential benefit of singular modes for multi-mode applications such as high-dimensional information encoding in dynamical disordered media.

	\emph{Conclusion.---}
	The universality of our approach may open new venues of future research in various directions.
	For instance, the split-step method and the subsequent decomposition of the transmission matrix may be applied to any channel geometry and any turbulent medium, including non-Kolmogorov models or underwater channels.
	It would be particularly interesting to examine how our main results
	are transferred to different dynamic disordered media.
	Finally, recent experimental efforts have been dedicated to the spatiotemporal characterization of disorder in optical media, see e.g. the measurement of a time-gated transmission matrix in \cite{DevaudPRA2022}.
	We expect that combining these experimental advances with our theoretical
	tools could lead to new levels of control of light transmission in time-dependent disordered optical media.  
	
	\emph{Acknowledgements.---}
	The authors are thankful for the fruitful discussions with Szymon Gładysz. We also acknowledge financial support of the Studienstiftung des deutschen Volkes
	and support by the state of Baden-Württemberg through bwHPC.
	This work was partially funded by French ANR under COSMIC project (ANR-19-ASTR0020-01). V.S. and A.B. acknowledge partial funding and support through the Strategiefonds der Albert-Ludwigs-Universität Freiburg and the Georg H. Endress Stiftung.

\bibliography{singularmodes4}

\begin{thebibliography}{48}%
\makeatletter
\providecommand \@ifxundefined [1]{%
 \@ifx{#1\undefined}
}%
\providecommand \@ifnum [1]{%
 \ifnum #1\expandafter \@firstoftwo
 \else \expandafter \@secondoftwo
 \fi
}%
\providecommand \@ifx [1]{%
 \ifx #1\expandafter \@firstoftwo
 \else \expandafter \@secondoftwo
 \fi
}%
\providecommand \natexlab [1]{#1}%
\providecommand \enquote  [1]{``#1''}%
\providecommand \bibnamefont  [1]{#1}%
\providecommand \bibfnamefont [1]{#1}%
\providecommand \citenamefont [1]{#1}%
\providecommand \href@noop [0]{\@secondoftwo}%
\providecommand \href [0]{\begingroup \@sanitize@url \@href}%
\providecommand \@href[1]{\@@startlink{#1}\@@href}%
\providecommand \@@href[1]{\endgroup#1\@@endlink}%
\providecommand \@sanitize@url [0]{\catcode `\\12\catcode `\$12\catcode
  `\&12\catcode `\#12\catcode `\^12\catcode `\_12\catcode `\%12\relax}%
\providecommand \@@startlink[1]{}%
\providecommand \@@endlink[0]{}%
\providecommand \url  [0]{\begingroup\@sanitize@url \@url }%
\providecommand \@url [1]{\endgroup\@href {#1}{\urlprefix }}%
\providecommand \urlprefix  [0]{URL }%
\providecommand \Eprint [0]{\href }%
\providecommand \doibase [0]{http://dx.doi.org/}%
\providecommand \selectlanguage [0]{\@gobble}%
\providecommand \bibinfo  [0]{\@secondoftwo}%
\providecommand \bibfield  [0]{\@secondoftwo}%
\providecommand \translation [1]{[#1]}%
\providecommand \BibitemOpen [0]{}%
\providecommand \bibitemStop [0]{}%
\providecommand \bibitemNoStop [0]{.\EOS\space}%
\providecommand \EOS [0]{\spacefactor3000\relax}%
\providecommand \BibitemShut  [1]{\csname bibitem#1\endcsname}%
\let\auto@bib@innerbib\@empty
\bibitem [{\citenamefont {\color{black}F. Roddier}(2004)}]{Roddier04}%
  \BibitemOpen
  \bibfield  {author} {\bibinfo {author} {\bibnamefont {\color{black}F.
  Roddier}},\ }\href
  {https://www.ebook.de/de/product/4208578/adaptive_optics_in_astronomy.html}
  {\emph {\bibinfo {title} {Adaptive Optics in Astronomy}}}\ (\bibinfo
  {publisher} {Cambridge University Press},\ \bibinfo {address} {Cambridge},\
  \bibinfo {year} {2004})\BibitemShut {NoStop}%
\bibitem [{\citenamefont {\color{black}G. Labeyrie}\ \emph
  {et~al.}(1999)\citenamefont {\color{black}G. Labeyrie}, \citenamefont
  {Tomasi}, \citenamefont {Bernard}, \citenamefont {M{\"u}ller}, \citenamefont
  {Miniatura},\ and\ \citenamefont {Kaiser}}]{Labeyrie99}%
  \BibitemOpen
  \bibfield  {author} {\bibinfo {author} {\bibnamefont {\color{black}G.
  Labeyrie}}, \bibinfo {author} {\bibfnamefont {F.}~\bibnamefont {Tomasi}},
  \bibinfo {author} {\bibfnamefont {J.}~\bibnamefont {Bernard}}, \bibinfo
  {author} {\bibfnamefont {C.~A.}\ \bibnamefont {M{\"u}ller}}, \bibinfo
  {author} {\bibfnamefont {C.}~\bibnamefont {Miniatura}}, \ and\ \bibinfo
  {author} {\bibfnamefont {R.}~\bibnamefont {Kaiser}},\ }\href {\doibase
  10.1103/physrevlett.83.5266} {\bibfield  {journal} {\bibinfo  {journal}
  {Phys. Rev. Lett.}\ }\textbf {\bibinfo {volume} {83}},\ \bibinfo {pages}
  {5266} (\bibinfo {year} {1999})}\BibitemShut {NoStop}%
\bibitem [{\citenamefont {Popoff}\ \emph {et~al.}(2014)\citenamefont {Popoff},
  \citenamefont {Goetschy}, \citenamefont {Liew}, \citenamefont {Stone},\ and\
  \citenamefont {Cao}}]{Popoff14}%
  \BibitemOpen
  \bibfield  {author} {\bibinfo {author} {\bibfnamefont {S.~M.}\ \bibnamefont
  {Popoff}}, \bibinfo {author} {\bibfnamefont {A.}~\bibnamefont {Goetschy}},
  \bibinfo {author} {\bibfnamefont {S.~F.}\ \bibnamefont {Liew}}, \bibinfo
  {author} {\bibfnamefont {A.~D.}\ \bibnamefont {Stone}}, \ and\ \bibinfo
  {author} {\bibfnamefont {H.}~\bibnamefont {Cao}},\ }\href {\doibase
  10.1103/PhysRevLett.112.133903} {\bibfield  {journal} {\bibinfo  {journal}
  {Phys. Rev. Lett.}\ }\textbf {\bibinfo {volume} {112}},\ \bibinfo {pages}
  {133903} (\bibinfo {year} {2014})}\BibitemShut {NoStop}%
\bibitem [{\citenamefont {Hsu}\ \emph {et~al.}(2017)\citenamefont {Hsu},
  \citenamefont {Liew}, \citenamefont {Goetschy}, \citenamefont {Cao},\ and\
  \citenamefont {Stone}}]{Hsu17}%
  \BibitemOpen
  \bibfield  {author} {\bibinfo {author} {\bibfnamefont {C.~W.}\ \bibnamefont
  {Hsu}}, \bibinfo {author} {\bibfnamefont {S.~F.}\ \bibnamefont {Liew}},
  \bibinfo {author} {\bibfnamefont {A.}~\bibnamefont {Goetschy}}, \bibinfo
  {author} {\bibfnamefont {H.}~\bibnamefont {Cao}}, \ and\ \bibinfo {author}
  {\bibfnamefont {A.~D.}\ \bibnamefont {Stone}},\ }\href {\doibase
  10.1038/nphys4036} {\bibfield  {journal} {\bibinfo  {journal} {Nature
  Physics}\ }\textbf {\bibinfo {volume} {13}},\ \bibinfo {pages} {497}
  (\bibinfo {year} {2017})}\BibitemShut {NoStop}%
\bibitem [{\citenamefont {Dorokhov}(1984)}]{DOROKHOV1984}%
  \BibitemOpen
  \bibfield  {author} {\bibinfo {author} {\bibfnamefont {O.~N.}\ \bibnamefont
  {Dorokhov}},\ }\href {\doibase https://doi.org/10.1016/0038-1098(84)90117-0}
  {\bibfield  {journal} {\bibinfo  {journal} {Solid State Commun.}\ }\textbf
  {\bibinfo {volume} {51}},\ \bibinfo {pages} {381} (\bibinfo {year}
  {1984})}\BibitemShut {NoStop}%
\bibitem [{\citenamefont {\color{black}Michael C.~Roggemann}\ \emph
  {et~al.}(1997)\citenamefont {\color{black}Michael C.~Roggemann},
  \citenamefont {Welsh},\ and\ \citenamefont {Fugate}}]{Roggemann97}%
  \BibitemOpen
  \bibfield  {author} {\bibinfo {author} {\bibnamefont {\color{black}Michael
  C.~Roggemann}}, \bibinfo {author} {\bibfnamefont {B.~M.}\ \bibnamefont
  {Welsh}}, \ and\ \bibinfo {author} {\bibfnamefont {R.~Q.}\ \bibnamefont
  {Fugate}},\ }\href {\doibase 10.1103/revmodphys.69.437} {\bibfield  {journal}
  {\bibinfo  {journal} {Rev. Mod. Phys.}\ }\textbf {\bibinfo {volume} {69}},\
  \bibinfo {pages} {437} (\bibinfo {year} {1997})}\BibitemShut {NoStop}%
\bibitem [{\citenamefont {Fante}(1975)}]{Fante75}%
  \BibitemOpen
  \bibfield  {author} {\bibinfo {author} {\bibfnamefont {R.}~\bibnamefont
  {Fante}},\ }\href {\doibase 10.1109/proc.1975.10035} {\bibfield  {journal}
  {\bibinfo  {journal} {Proceedings of the {IEEE}}\ }\textbf {\bibinfo {volume}
  {63}},\ \bibinfo {pages} {1669} (\bibinfo {year} {1975})}\BibitemShut
  {NoStop}%
\bibitem [{\citenamefont {\color{black}Y. Ren}\ \emph
  {et~al.}(2014)\citenamefont {\color{black}Y. Ren}, \citenamefont {Xie},
  \citenamefont {Huang}, \citenamefont {Bao}, \citenamefont {Yan},
  \citenamefont {Ahmed}, \citenamefont {Lavery}, \citenamefont {Erkmen},
  \citenamefont {Dolinar}, \citenamefont {Tur}, \citenamefont {Neifeld},
  \citenamefont {Padgett}, \citenamefont {Boyd}, \citenamefont {Shapiro},\ and\
  \citenamefont {Willner}}]{Ren14}%
  \BibitemOpen
  \bibfield  {author} {\bibinfo {author} {\bibnamefont {\color{black}Y. Ren}},
  \bibinfo {author} {\bibfnamefont {G.}~\bibnamefont {Xie}}, \bibinfo {author}
  {\bibfnamefont {H.}~\bibnamefont {Huang}}, \bibinfo {author} {\bibfnamefont
  {C.}~\bibnamefont {Bao}}, \bibinfo {author} {\bibfnamefont {Y.}~\bibnamefont
  {Yan}}, \bibinfo {author} {\bibfnamefont {N.}~\bibnamefont {Ahmed}}, \bibinfo
  {author} {\bibfnamefont {M.~P.~J.}\ \bibnamefont {Lavery}}, \bibinfo {author}
  {\bibfnamefont {B.~I.}\ \bibnamefont {Erkmen}}, \bibinfo {author}
  {\bibfnamefont {S.}~\bibnamefont {Dolinar}}, \bibinfo {author} {\bibfnamefont
  {M.}~\bibnamefont {Tur}}, \bibinfo {author} {\bibfnamefont {M.~A.}\
  \bibnamefont {Neifeld}}, \bibinfo {author} {\bibfnamefont {M.~J.}\
  \bibnamefont {Padgett}}, \bibinfo {author} {\bibfnamefont {R.~W.}\
  \bibnamefont {Boyd}}, \bibinfo {author} {\bibfnamefont {J.~H.}\ \bibnamefont
  {Shapiro}}, \ and\ \bibinfo {author} {\bibfnamefont {A.~E.}\ \bibnamefont
  {Willner}},\ }\href {\doibase 10.1364/ol.39.002845} {\bibfield  {journal}
  {\bibinfo  {journal} {Optics Lett.}\ }\textbf {\bibinfo {volume} {39}},\
  \bibinfo {pages} {2845} (\bibinfo {year} {2014})}\BibitemShut {NoStop}%
\bibitem [{\citenamefont {\color{black}Y. Ren}\ \emph
  {et~al.}(2016)\citenamefont {\color{black}Y. Ren}, \citenamefont {Li},
  \citenamefont {Wang}, \citenamefont {Kamali}, \citenamefont {Arbabi},
  \citenamefont {Arbabi}, \citenamefont {Zhao}, \citenamefont {Xie},
  \citenamefont {Cao}, \citenamefont {Ahmed}, \citenamefont {Yan},
  \citenamefont {Liu}, \citenamefont {Willner}, \citenamefont {Ashrafi},
  \citenamefont {Tur}, \citenamefont {Faraon},\ and\ \citenamefont
  {Willner}}]{Ren16}%
  \BibitemOpen
  \bibfield  {author} {\bibinfo {author} {\bibnamefont {\color{black}Y. Ren}},
  \bibinfo {author} {\bibfnamefont {L.}~\bibnamefont {Li}}, \bibinfo {author}
  {\bibfnamefont {Z.}~\bibnamefont {Wang}}, \bibinfo {author} {\bibfnamefont
  {S.~M.}\ \bibnamefont {Kamali}}, \bibinfo {author} {\bibfnamefont
  {E.}~\bibnamefont {Arbabi}}, \bibinfo {author} {\bibfnamefont
  {A.}~\bibnamefont {Arbabi}}, \bibinfo {author} {\bibfnamefont
  {Z.}~\bibnamefont {Zhao}}, \bibinfo {author} {\bibfnamefont {G.}~\bibnamefont
  {Xie}}, \bibinfo {author} {\bibfnamefont {Y.}~\bibnamefont {Cao}}, \bibinfo
  {author} {\bibfnamefont {N.}~\bibnamefont {Ahmed}}, \bibinfo {author}
  {\bibfnamefont {Y.}~\bibnamefont {Yan}}, \bibinfo {author} {\bibfnamefont
  {C.}~\bibnamefont {Liu}}, \bibinfo {author} {\bibfnamefont {A.~J.}\
  \bibnamefont {Willner}}, \bibinfo {author} {\bibfnamefont {S.}~\bibnamefont
  {Ashrafi}}, \bibinfo {author} {\bibfnamefont {M.}~\bibnamefont {Tur}},
  \bibinfo {author} {\bibfnamefont {A.}~\bibnamefont {Faraon}}, \ and\ \bibinfo
  {author} {\bibfnamefont {A.~E.}\ \bibnamefont {Willner}},\ }\href
  {https://doi.org/10.1038%2Fsrep33306} {\bibfield  {journal} {\bibinfo
  {journal} {Sci. Rep.}\ }\textbf {\bibinfo {volume} {6}},\ \bibinfo {pages}
  {33306} (\bibinfo {year} {2016})}\BibitemShut {NoStop}%
\bibitem [{\citenamefont {\color{black}J. Zhao}\ \emph
  {et~al.}(2020)\citenamefont {\color{black}J. Zhao}, \citenamefont {Zhou},
  \citenamefont {Braverman}, \citenamefont {Liu}, \citenamefont {Pang},
  \citenamefont {Steinhoff}, \citenamefont {Tyler}, \citenamefont {Willner},\
  and\ \citenamefont {Boyd}}]{Zhao20}%
  \BibitemOpen
  \bibfield  {author} {\bibinfo {author} {\bibnamefont {\color{black}J. Zhao}},
  \bibinfo {author} {\bibfnamefont {Y.}~\bibnamefont {Zhou}}, \bibinfo {author}
  {\bibfnamefont {B.}~\bibnamefont {Braverman}}, \bibinfo {author}
  {\bibfnamefont {C.}~\bibnamefont {Liu}}, \bibinfo {author} {\bibfnamefont
  {K.}~\bibnamefont {Pang}}, \bibinfo {author} {\bibfnamefont {N.~K.}\
  \bibnamefont {Steinhoff}}, \bibinfo {author} {\bibfnamefont {G.~A.}\
  \bibnamefont {Tyler}}, \bibinfo {author} {\bibfnamefont {A.~E.}\ \bibnamefont
  {Willner}}, \ and\ \bibinfo {author} {\bibfnamefont {R.~W.}\ \bibnamefont
  {Boyd}},\ }\href {\doibase 10.1364/oe.390518} {\bibfield  {journal} {\bibinfo
   {journal} {Opt. Express}\ }\textbf {\bibinfo {volume} {28}},\ \bibinfo
  {pages} {15376} (\bibinfo {year} {2020})}\BibitemShut {NoStop}%
\bibitem [{\citenamefont {\color{black}Tyson}(2016)}]{Tyson16}%
  \BibitemOpen
  \bibfield  {author} {\bibinfo {author} {\bibfnamefont {R.}~\bibnamefont
  {\color{black}Tyson}},\ }\href@noop {} {\emph {\bibinfo {title} {Principles
  of adaptive optics}}}\ (\bibinfo  {publisher} {CRC Press},\ \bibinfo
  {address} {Boca Raton},\ \bibinfo {year} {2016})\BibitemShut {NoStop}%
\bibitem [{\citenamefont {\color{black}G. Sorelli}\ \emph
  {et~al.}(2019)\citenamefont {\color{black}G. Sorelli}, \citenamefont
  {Leonhard}, \citenamefont {Shatokhin}, \citenamefont {Reinlein},\ and\
  \citenamefont {Buchleitner}}]{Sorelli19}%
  \BibitemOpen
  \bibfield  {author} {\bibinfo {author} {\bibnamefont {\color{black}G.
  Sorelli}}, \bibinfo {author} {\bibfnamefont {N.}~\bibnamefont {Leonhard}},
  \bibinfo {author} {\bibfnamefont {V.~N.}\ \bibnamefont {Shatokhin}}, \bibinfo
  {author} {\bibfnamefont {C.}~\bibnamefont {Reinlein}}, \ and\ \bibinfo
  {author} {\bibfnamefont {A.}~\bibnamefont {Buchleitner}},\ }\href {\doibase
  10.1088/1367-2630/ab006e} {\bibfield  {journal} {\bibinfo  {journal} {New J.
  Phys.}\ }\textbf {\bibinfo {volume} {21}},\ \bibinfo {pages} {023003}
  (\bibinfo {year} {2019})}\BibitemShut {NoStop}%
\bibitem [{\citenamefont {\color{black}S. Rotter}\ and\ \citenamefont
  {Gigan}(2017)}]{Rotter17}%
  \BibitemOpen
  \bibfield  {author} {\bibinfo {author} {\bibnamefont {\color{black}S.
  Rotter}}\ and\ \bibinfo {author} {\bibfnamefont {S.}~\bibnamefont {Gigan}},\
  }\href {https://doi.org/10.1103%2Frevmodphys.89.015005} {\bibfield  {journal}
  {\bibinfo  {journal} {Reviews of Modern Physics}\ }\textbf {\bibinfo {volume}
  {89}},\ \bibinfo {pages} {015005} (\bibinfo {year} {2017})}\BibitemShut
  {NoStop}%
\bibitem [{\citenamefont {Segev}\ \emph {et~al.}(2013)\citenamefont {Segev},
  \citenamefont {Silberberg},\ and\ \citenamefont {Christodoulides}}]{Segev13}%
  \BibitemOpen
  \bibfield  {author} {\bibinfo {author} {\bibfnamefont {M.}~\bibnamefont
  {Segev}}, \bibinfo {author} {\bibfnamefont {Y.}~\bibnamefont {Silberberg}}, \
  and\ \bibinfo {author} {\bibfnamefont {D.~N.}\ \bibnamefont
  {Christodoulides}},\ }\href {\doibase 10.1038/nphoton.2013.30} {\bibfield
  {journal} {\bibinfo  {journal} {Nat. Photonics}\ }\textbf {\bibinfo {volume}
  {7}},\ \bibinfo {pages} {197} (\bibinfo {year} {2013})}\BibitemShut {NoStop}%
\bibitem [{\citenamefont {Y{\i}lmaz}\ \emph {et~al.}(2019)\citenamefont
  {Y{\i}lmaz}, \citenamefont {Hsu}, \citenamefont {Yamilov},\ and\
  \citenamefont {Cao}}]{Yi19}%
  \BibitemOpen
  \bibfield  {author} {\bibinfo {author} {\bibfnamefont {H.}~\bibnamefont
  {Y{\i}lmaz}}, \bibinfo {author} {\bibfnamefont {C.~W.}\ \bibnamefont {Hsu}},
  \bibinfo {author} {\bibfnamefont {A.}~\bibnamefont {Yamilov}}, \ and\
  \bibinfo {author} {\bibfnamefont {H.}~\bibnamefont {Cao}},\ }\href {\doibase
  10.1038/s41566-019-0367-9} {\bibfield  {journal} {\bibinfo  {journal} {Nature
  Photonics}\ }\textbf {\bibinfo {volume} {13}},\ \bibinfo {pages} {352}
  (\bibinfo {year} {2019})}\BibitemShut {NoStop}%
\bibitem [{\citenamefont {Kopeika}(1998)}]{Kopeika98}%
  \BibitemOpen
  \bibfield  {author} {\bibinfo {author} {\bibfnamefont {N.}~\bibnamefont
  {Kopeika}},\ }\href@noop {} {\emph {\bibinfo {title} {A System Engineering
  Approach to Imaging}}},\ Press Monographs\ (\bibinfo  {publisher} {SPIE
  Optical Engineering Press},\ \bibinfo {year} {1998})\BibitemShut {NoStop}%
\bibitem [{\citenamefont {\color{black}Andrews}(2005)}]{Andrews05}%
  \BibitemOpen
  \bibfield  {author} {\bibinfo {author} {\bibfnamefont {L.}~\bibnamefont
  {\color{black}Andrews}},\ }\href@noop {} {\emph {\bibinfo {title} {Laser beam
  propagation through random media}}}\ (\bibinfo  {publisher} {SPIE},\ \bibinfo
  {address} {Bellingham},\ \bibinfo {year} {2005})\BibitemShut {NoStop}%
\bibitem [{\citenamefont {Carminati}\ and\ \citenamefont
  {Schotland}(2021)}]{Carminati21}%
  \BibitemOpen
  \bibfield  {author} {\bibinfo {author} {\bibfnamefont {R.}~\bibnamefont
  {Carminati}}\ and\ \bibinfo {author} {\bibfnamefont {J.~C.}\ \bibnamefont
  {Schotland}},\ }\href {\doibase 10.1017/9781316544693} {\emph {\bibinfo
  {title} {Principles of Scattering and Transport of Light}}}\ (\bibinfo
  {publisher} {Cambridge University Press},\ \bibinfo {year}
  {2021})\BibitemShut {NoStop}%
\bibitem [{mea()}]{meanfreepath}%
  \BibitemOpen
  \href@noop {} {\bibinfo  {journal} {For the propagation distance of $1\;$km,
  the power transmittance at $\lambda=1550\,$nm is about 99\% \cite{Andrews05},
  which via the Lambert-Beer law yields the transport mean free path $\simeq
  100\,$km}\ }\BibitemShut {NoStop}%
\bibitem [{\citenamefont {Kolmogorov}(1941)}]{Kolmogorov41a}%
  \BibitemOpen
\bibfield  {journal} {  }\bibfield  {author} {\bibinfo {author} {\bibfnamefont
  {A.~N.}\ \bibnamefont {Kolmogorov}},\ }\href@noop {} {\bibfield  {journal}
  {\bibinfo  {journal} {Dokl. Acad. Sci. URSS}\ }\textbf {\bibinfo {volume}
  {30}},\ \bibinfo {pages} {301} (\bibinfo {year} {1941})}\BibitemShut
  {NoStop}%
\bibitem [{\citenamefont {\color{black}Tatarskii}(2016)}]{Tatarskii16}%
  \BibitemOpen
  \bibfield  {author} {\bibinfo {author} {\bibfnamefont {V.~I.}\ \bibnamefont
  {\color{black}Tatarskii}},\ }\href@noop {} {\emph {\bibinfo {title} {Wave
  propagation in a turbulent medium}}}\ (\bibinfo  {publisher} {Dover
  Publications},\ \bibinfo {address} {Mineola},\ \bibinfo {year}
  {2016})\BibitemShut {NoStop}%
\bibitem [{\citenamefont {\color{black}A. Ishimaru}(1978)}]{Ishimaru78}%
  \BibitemOpen
  \bibfield  {author} {\bibinfo {author} {\bibnamefont {\color{black}A.
  Ishimaru}},\ }\href {\doibase 10.1016/b978-0-12-374701-3.x5001-7} {\emph
  {\bibinfo {title} {Wave Propagation and Scattering in Random Media}}}\
  (\bibinfo  {publisher} {Academic Press},\ \bibinfo {address} {New York},\
  \bibinfo {year} {1978})\BibitemShut {NoStop}%
\bibitem [{\citenamefont {\color{black}Schmidt}(2010)}]{Schmidt10}%
  \BibitemOpen
  \bibfield  {author} {\bibinfo {author} {\bibfnamefont {J.}~\bibnamefont
  {\color{black}Schmidt}},\ }\href@noop {} {\emph {\bibinfo {title} {Numerical
  simulation of optical wave propagation with examples in MATLAB}}}\ (\bibinfo
  {publisher} {SPIE},\ \bibinfo {address} {Bellingham},\ \bibinfo {year}
  {2010})\BibitemShut {NoStop}%
\bibitem [{\citenamefont {\color{black}Lukin}(2002)}]{Lukin02}%
  \BibitemOpen
  \bibfield  {author} {\bibinfo {author} {\bibfnamefont {V.~P.}\ \bibnamefont
  {\color{black}Lukin}},\ }\href@noop {} {\emph {\bibinfo {title} {Adaptive
  beaming and imaging in the turbulent atmosphere}}}\ (\bibinfo  {publisher}
  {SPIE},\ \bibinfo {address} {Bellingham},\ \bibinfo {year}
  {2002})\BibitemShut {NoStop}%
\bibitem [{sup()}]{supp}%
  \BibitemOpen
  \href@noop {} {\bibinfo  {journal} {See Supplemental Material for more
  details of the mean transmitted power $P_{s}(t/t_\text{c})$ of the first
  eleven singular modes, the fitting procedures and the Gaussian (i.e.
  non-Kolmogorov) power spectrum}\ }\BibitemShut {NoStop}%
\bibitem [{\citenamefont {\color{black}Born}\ and\ \citenamefont
  {Wolf}(1999)}]{Born99}%
  \BibitemOpen
\bibfield  {journal} {  }\bibfield  {author} {\bibinfo {author} {\bibfnamefont
  {M.}~\bibnamefont {\color{black}Born}}\ and\ \bibinfo {author} {\bibfnamefont
  {E.}~\bibnamefont {Wolf}},\ }\href@noop {} {\emph {\bibinfo {title}
  {Principles of optics : electromagnetic theory of propagation, interference
  and diffraction of light}}}\ (\bibinfo  {publisher} {Cambridge University
  Press},\ \bibinfo {address} {Cambridge},\ \bibinfo {year} {1999})\BibitemShut
  {NoStop}%
\bibitem [{\citenamefont {\color{black}Goodman}(2005)}]{Goodman05}%
  \BibitemOpen
  \bibfield  {author} {\bibinfo {author} {\bibfnamefont {J.}~\bibnamefont
  {\color{black}Goodman}},\ }\href@noop {} {\emph {\bibinfo {title}
  {Introduction to Fourier optics}}}\ (\bibinfo  {publisher} {Roberts \& Co},\
  \bibinfo {address} {Englewood},\ \bibinfo {year} {2005})\BibitemShut
  {NoStop}%
\bibitem [{\citenamefont {\color{black}E. M.~Johansson}\ and\ \citenamefont
  {Gavel}(1994)}]{Johansson94}%
  \BibitemOpen
  \bibfield  {author} {\bibinfo {author} {\bibnamefont {\color{black}E.
  M.~Johansson}}\ and\ \bibinfo {author} {\bibfnamefont {D.~T.}\ \bibnamefont
  {Gavel}},\ }in\ \href {\doibase 10.1117/12.177254} {\emph {\bibinfo
  {booktitle} {Amplitude and Intensity Spatial Interferometry {II}}}},\
  \bibinfo {editor} {edited by\ \bibinfo {editor} {\bibfnamefont {J.~B.}\
  \bibnamefont {Breckinridge}}}\ (\bibinfo  {publisher} {{SPIE}},\ \bibinfo
  {year} {1994})\BibitemShut {NoStop}%
\bibitem [{\citenamefont {\color{black}N. A.~Roddier}(1990)}]{Roddier90}%
  \BibitemOpen
  \bibfield  {author} {\bibinfo {author} {\bibnamefont {\color{black}N.
  A.~Roddier}},\ }\href {\doibase 10.1117/12.55712} {\bibfield  {journal}
  {\bibinfo  {journal} {Opt. Eng.}\ }\textbf {\bibinfo {volume} {29}},\
  \bibinfo {pages} {1174} (\bibinfo {year} {1990})}\BibitemShut {NoStop}%
\bibitem [{\citenamefont {\color{black}D. A.~B.~Miller}(2019)}]{Miller19}%
  \BibitemOpen
  \bibfield  {author} {\bibinfo {author} {\bibnamefont {\color{black}D.
  A.~B.~Miller}},\ }\href {\doibase 10.1364/aop.11.000679} {\bibfield
  {journal} {\bibinfo  {journal} {Adv. Opt. Photonics}\ }\textbf {\bibinfo
  {volume} {11}},\ \bibinfo {pages} {679} (\bibinfo {year} {2019})}\BibitemShut
  {NoStop}%
\bibitem [{our(lack{\natexlab{a}})}]{ourchannel}%
  \BibitemOpen
  \href@noop {} {\bibfield  {journal} {\bibinfo  {journal}
  {\color{black}Parameters of our channel: $L=2.5\,$km, $D_\text{s}=20\,$cm,
  $D_\text{r}=25\,$cm. Source basis: $S=496$ LG modes (beam waist $18\,$mm,
  radial and azimuthal indices $p,|\ell| \le 15$); receiver basis: $Q=221^2$
  (pixels \color{white}}\ } (\bibinfo {year}
  {\hspace{-13pt}\color{black}}{\natexlab{a}})}\BibitemShut {NoStop}%
\bibitem [{\citenamefont {Lavery}\ \emph {et~al.}(2017)\citenamefont {Lavery},
  \citenamefont {Peuntinger}, \citenamefont {Günthner}, \citenamefont
  {Banzer}, \citenamefont {Elser}, \citenamefont {Boyd}, \citenamefont
  {Padgett}, \citenamefont {Marquardt},\ and\ \citenamefont
  {Leuchs}}]{Lavery17}%
  \BibitemOpen
  \bibfield  {author} {\bibinfo {author} {\bibfnamefont {M.~P.~J.}\
  \bibnamefont {Lavery}}, \bibinfo {author} {\bibfnamefont {C.}~\bibnamefont
  {Peuntinger}}, \bibinfo {author} {\bibfnamefont {K.}~\bibnamefont
  {Günthner}}, \bibinfo {author} {\bibfnamefont {P.}~\bibnamefont {Banzer}},
  \bibinfo {author} {\bibfnamefont {D.}~\bibnamefont {Elser}}, \bibinfo
  {author} {\bibfnamefont {R.~W.}\ \bibnamefont {Boyd}}, \bibinfo {author}
  {\bibfnamefont {M.~J.}\ \bibnamefont {Padgett}}, \bibinfo {author}
  {\bibfnamefont {C.}~\bibnamefont {Marquardt}}, \ and\ \bibinfo {author}
  {\bibfnamefont {G.}~\bibnamefont {Leuchs}},\ }\href {\doibase
  10.1126/sciadv.1700552} {\bibfield  {journal} {\bibinfo  {journal} {Science
  Advances}\ }\textbf {\bibinfo {volume} {3}},\ \bibinfo {pages} {e1700552}
  (\bibinfo {year} {2017})}\BibitemShut {NoStop}%
\bibitem [{\citenamefont {\color{black}M. Krenn}\ \emph
  {et~al.}(2015)\citenamefont {\color{black}M. Krenn}, \citenamefont
  {Handsteiner}, \citenamefont {Fink}, \citenamefont {Fickler},\ and\
  \citenamefont {Zeilinger}}]{Krenn15}%
  \BibitemOpen
  \bibfield  {author} {\bibinfo {author} {\bibnamefont {\color{black}M.
  Krenn}}, \bibinfo {author} {\bibfnamefont {J.}~\bibnamefont {Handsteiner}},
  \bibinfo {author} {\bibfnamefont {M.}~\bibnamefont {Fink}}, \bibinfo {author}
  {\bibfnamefont {R.}~\bibnamefont {Fickler}}, \ and\ \bibinfo {author}
  {\bibfnamefont {A.}~\bibnamefont {Zeilinger}},\ }\href {\doibase
  10.1073/pnas.1517574112} {\bibfield  {journal} {\bibinfo  {journal} {Proc.
  Natl. Acad. Sci. U.S.A.}\ }\textbf {\bibinfo {volume} {112}},\ \bibinfo
  {pages} {14197} (\bibinfo {year} {2015})}\BibitemShut {NoStop}%
\bibitem [{\citenamefont {\color{black}J. H.~Shapiro}(1974)}]{Shapiro74}%
  \BibitemOpen
  \bibfield  {author} {\bibinfo {author} {\bibnamefont {\color{black}J.
  H.~Shapiro}},\ }\href {\doibase 10.1364/ao.13.002614} {\bibfield  {journal}
  {\bibinfo  {journal} {Appl. Opt.}\ }\textbf {\bibinfo {volume} {13}},\
  \bibinfo {pages} {2614} (\bibinfo {year} {1974})}\BibitemShut {NoStop}%
\bibitem [{\citenamefont {\color{black}J. M.~Kahn}\ and\ \citenamefont
  {Belmonte}(2018)}]{Kahn18}%
  \BibitemOpen
  \bibfield  {author} {\bibinfo {author} {\bibnamefont {\color{black}J.
  M.~Kahn}}\ and\ \bibinfo {author} {\bibfnamefont {A.}~\bibnamefont
  {Belmonte}},\ }in\ \href {\doibase 10.1117/12.2299623} {\emph {\bibinfo
  {booktitle} {Broadband Access Communication Technologies {XII}}}},\ \bibinfo
  {editor} {edited by\ \bibinfo {editor} {\bibfnamefont {B.~B.}\ \bibnamefont
  {Dingel}}, \bibinfo {editor} {\bibfnamefont {K.}~\bibnamefont {Tsukamoto}}, \
  and\ \bibinfo {editor} {\bibfnamefont {S.}~\bibnamefont {Mikroulis}}}\
  (\bibinfo  {publisher} {{SPIE}},\ \bibinfo {year} {2018})\BibitemShut
  {NoStop}%
\bibitem [{\citenamefont {\color{black}L. Borcea}\ \emph
  {et~al.}(2020)\citenamefont {\color{black}L. Borcea}, \citenamefont
  {Garnier},\ and\ \citenamefont {S{\o}lna}}]{Borcea20}%
  \BibitemOpen
  \bibfield  {author} {\bibinfo {author} {\bibnamefont {\color{black}L.
  Borcea}}, \bibinfo {author} {\bibfnamefont {J.}~\bibnamefont {Garnier}}, \
  and\ \bibinfo {author} {\bibfnamefont {K.}~\bibnamefont {S{\o}lna}},\ }\href
  {\doibase 10.1364/josaa.384007} {\bibfield  {journal} {\bibinfo  {journal}
  {J. Opt. Soc. Am. A}\ }\textbf {\bibinfo {volume} {37}},\ \bibinfo {pages}
  {720} (\bibinfo {year} {2020})}\BibitemShut {NoStop}%
\bibitem [{\citenamefont {\color{black}V. N.~Shatokhin}\ \emph
  {et~al.}(2020)\citenamefont {\color{black}V. N.~Shatokhin}, \citenamefont
  {Bachmann}, \citenamefont {Sorelli}, \citenamefont {Treps},\ and\
  \citenamefont {Buchleitner}}]{Shatokhin20}%
  \BibitemOpen
  \bibfield  {author} {\bibinfo {author} {\bibnamefont {\color{black}V.
  N.~Shatokhin}}, \bibinfo {author} {\bibfnamefont {D.}~\bibnamefont
  {Bachmann}}, \bibinfo {author} {\bibfnamefont {G.}~\bibnamefont {Sorelli}},
  \bibinfo {author} {\bibfnamefont {N.}~\bibnamefont {Treps}}, \ and\ \bibinfo
  {author} {\bibfnamefont {A.}~\bibnamefont {Buchleitner}},\ }in\ \href
  {\doibase 10.1117/12.2573477} {\emph {\bibinfo {booktitle} {Environmental
  Effects on Light Propagation and Adaptive Systems {III}}}},\ \bibinfo
  {editor} {edited by\ \bibinfo {editor} {\bibfnamefont {K.}~\bibnamefont
  {Stein}}\ and\ \bibinfo {editor} {\bibfnamefont {S.}~\bibnamefont
  {Gladysz}}}\ (\bibinfo  {publisher} {{SPIE}},\ \bibinfo {year}
  {2020})\BibitemShut {NoStop}%
\bibitem [{\citenamefont {\color{black}D. Slepian}(1965)}]{Slepian65}%
  \BibitemOpen
  \bibfield  {author} {\bibinfo {author} {\bibnamefont {\color{black}D.
  Slepian}},\ }\href {\doibase 10.1364/josa.55.001110} {\bibfield  {journal}
  {\bibinfo  {journal} {J. Opt. Soc. Am.}\ }\textbf {\bibinfo {volume} {55}},\
  \bibinfo {pages} {1110} (\bibinfo {year} {1965})}\BibitemShut {NoStop}%
\bibitem [{\citenamefont {Nazarov}(1994)}]{NazarovPRL1994}%
  \BibitemOpen
  \bibfield  {author} {\bibinfo {author} {\bibfnamefont {Y.~V.}\ \bibnamefont
  {Nazarov}},\ }\href {\doibase 10.1103/PhysRevLett.73.134} {\bibfield
  {journal} {\bibinfo  {journal} {Phys. Rev. Lett.}\ }\textbf {\bibinfo
  {volume} {73}},\ \bibinfo {pages} {134} (\bibinfo {year} {1994})}\BibitemShut
  {NoStop}%
\bibitem [{\citenamefont {Beenakker}(1997)}]{BennakerRevModPhys}%
  \BibitemOpen
  \bibfield  {author} {\bibinfo {author} {\bibfnamefont {C.~W.~J.}\
  \bibnamefont {Beenakker}},\ }\href {\doibase 10.1103/RevModPhys.69.731}
  {\bibfield  {journal} {\bibinfo  {journal} {Rev. Mod. Phys.}\ }\textbf
  {\bibinfo {volume} {69}},\ \bibinfo {pages} {731} (\bibinfo {year}
  {1997})}\BibitemShut {NoStop}%
\bibitem [{\citenamefont {Goetschy}\ and\ \citenamefont
  {Stone}(2013)}]{GoetschyPRL2013}%
  \BibitemOpen
  \bibfield  {author} {\bibinfo {author} {\bibfnamefont {A.}~\bibnamefont
  {Goetschy}}\ and\ \bibinfo {author} {\bibfnamefont {A.~D.}\ \bibnamefont
  {Stone}},\ }\href {\doibase 10.1103/PhysRevLett.111.063901} {\bibfield
  {journal} {\bibinfo  {journal} {Phys. Rev. Lett.}\ }\textbf {\bibinfo
  {volume} {111}},\ \bibinfo {pages} {063901} (\bibinfo {year}
  {2013})}\BibitemShut {NoStop}%
\bibitem [{\citenamefont {\color{black}G. I.~Taylor}(1935)}]{Taylor35}%
  \BibitemOpen
  \bibfield  {author} {\bibinfo {author} {\bibnamefont {\color{black}G.
  I.~Taylor}},\ }\href {\doibase 10.1098/rspa.1935.0158} {\bibfield  {journal}
  {\bibinfo  {journal} {Proc. R. Soc. Lond. A}\ }\textbf {\bibinfo {volume}
  {151}},\ \bibinfo {pages} {421} (\bibinfo {year} {1935})}\BibitemShut
  {NoStop}%
\bibitem [{\citenamefont {\color{black}E. Anzuola}\ and\ \citenamefont
  {Gładysz}(2017)}]{Anzuola17}%
  \BibitemOpen
  \bibfield  {author} {\bibinfo {author} {\bibnamefont {\color{black}E.
  Anzuola}}\ and\ \bibinfo {author} {\bibfnamefont {S.}~\bibnamefont
  {Gładysz}},\ }\href {\doibase 10.1117/1.oe.56.7.071508} {\bibfield
  {journal} {\bibinfo  {journal} {Opt. Eng.}\ }\textbf {\bibinfo {volume}
  {56}},\ \bibinfo {pages} {1} (\bibinfo {year} {2017})}\BibitemShut {NoStop}%
\bibitem [{Note1()}]{Note1}%
  \BibitemOpen
  \bibinfo {note} {The LG beam waist $w_0=29.7\protect \,$mm is chosen to
  maximize transmission in vacuum. For the definitions of transmitted power
  $P_s(t)$ and crosstalk $C_{s',s}(t)$ of LG modes $s$ denotes the azimuthal
  index $\ell =0,\pm 1,\protect \dots ,\pm 5$ and $f_{s'}(\protect \boldsymbol
  {\rho }^\prime ,t)$ is given by modes propagated through evolved turbulence
  at time $t$ while $u_s(\protect \boldsymbol {\rho }^\prime )$ corresponds to
  LG modes propagated through vacuum.}\BibitemShut {Stop}%
\bibitem [{our(lack{\natexlab{b}})}]{ourcrosstalk}%
  \BibitemOpen
  \href@noop {} {\bibfield  {journal} {\bibinfo  {journal}
  {\color{black}Crosstalk matrices are inset in Fig.~\ref{fig:power} at
  $t/t_\text{c} = 0.95, \, 6.36, \, 14.00$ and 21.63. The entries correspond to
  eleven singular modes (top) or $p=0,\,|l|\leq 5$ optimized LG modes (bottom),
  where the vertical axes represent unperturbed modes at $t=0$ and transmitted
  modes run along the horizontal axes (ordered by transmittance \color{white}}\
  } (\bibinfo {year} {\hspace{-12pt}\color{black}}{\natexlab{b}})}\BibitemShut
  {NoStop}%
\bibitem [{our(lack{\natexlab{c}})}]{ourwind}%
  \BibitemOpen
  \href@noop {} {\bibfield  {journal} {\bibinfo  {journal}
  {\color{black}$\langle V \rangle=3\,$m/s and $\Delta V=1\,$m/s (Gaussian
  distribution \color{white}}\ } (\bibinfo {year}
  {\hspace{-12pt}\color{black}}{\natexlab{c}})}\BibitemShut {NoStop}%
\bibitem [{\citenamefont {\color{black}J. M.~Martin}\ and\ \citenamefont
  {Flatt{\'{e}}}(1988)}]{Martin88}%
  \BibitemOpen
  \bibfield  {author} {\bibinfo {author} {\bibnamefont {\color{black}J.
  M.~Martin}}\ and\ \bibinfo {author} {\bibfnamefont {S.~M.}\ \bibnamefont
  {Flatt{\'{e}}}},\ }\href {\doibase 10.1364/ao.27.002111} {\bibfield
  {journal} {\bibinfo  {journal} {Appl. Opt.}\ }\textbf {\bibinfo {volume}
  {27}},\ \bibinfo {pages} {2111} (\bibinfo {year} {1988})}\BibitemShut
  {NoStop}%
\bibitem [{\citenamefont {Devaud}\ \emph {et~al.}(2022)\citenamefont {Devaud},
  \citenamefont {Rauer}, \citenamefont {K\"uhmayer}, \citenamefont {Melchard},
  \citenamefont {Mounaix}, \citenamefont {Rotter},\ and\ \citenamefont
  {Gigan}}]{DevaudPRA2022}%
  \BibitemOpen
  \bibfield  {author} {\bibinfo {author} {\bibfnamefont {L.}~\bibnamefont
  {Devaud}}, \bibinfo {author} {\bibfnamefont {B.}~\bibnamefont {Rauer}},
  \bibinfo {author} {\bibfnamefont {M.}~\bibnamefont {K\"uhmayer}}, \bibinfo
  {author} {\bibfnamefont {J.}~\bibnamefont {Melchard}}, \bibinfo {author}
  {\bibfnamefont {M.}~\bibnamefont {Mounaix}}, \bibinfo {author} {\bibfnamefont
  {S.}~\bibnamefont {Rotter}}, \ and\ \bibinfo {author} {\bibfnamefont
  {S.}~\bibnamefont {Gigan}},\ }\href {\doibase 10.1103/PhysRevA.105.L051501}
  {\bibfield  {journal} {\bibinfo  {journal} {Phys. Rev. A}\ }\textbf {\bibinfo
  {volume} {105}},\ \bibinfo {pages} {L051501} (\bibinfo {year}
  {2022})}\BibitemShut {NoStop}%
\end{thebibliography}%

		\clearpage
		\section{Supplemental Material}
		\noindent In this Supplemental Material, we provide details of the mean transmitted power $P_{s}(t)$ (cf. Fig.~3 in the main manuscript) of the first eleven singular modes,
		i.e., $s=0,\ldots,10$, and for different values of the Fried parameter $r_0$, ranging from 15\,mm to 33\,mm (with a mean velocity $\langle V \rangle = 3$\,m/s). Table~\ref{supp:table1} summarizes the employed turbulence parameters. The number of phase screens $N_\text{screens}$ required for the split-step method at a given Fried parameter is chosen such that each interval is short enough to remain within the regime of weak intensity fluctuations, i.e., $\sigma_{\text{R},\text{step}}^2(r_0, \Delta z) \lesssim 1$.
		
		\begin{table}[h!]
			\centering
			\begin{tabular}{l|r|r|r|r}
				$r_0\,$/mm & 15 & 20 & 25 & 33 \\\hline\hline
				$\sigma_\text{R}^2$ & 6.72 & 4.16 & 2.87 & 1.81
				\\\hline
				$N_\text{screens}$ & 25 & 15 & 11 & 7 \\
				
			\end{tabular}
		\caption{Considered Fried parameters $r_0$ for atmospheric turbulence with corresponding Rytov variances $\sigma_\text{R}^2$ and required number of phase screens $N_\text{screens}$ of our numerical simulations.
		}
		\label{supp:table1}
	\end{table}
	
\begin{figure*}[h!]
		\includegraphics[width=0.8\columnwidth]{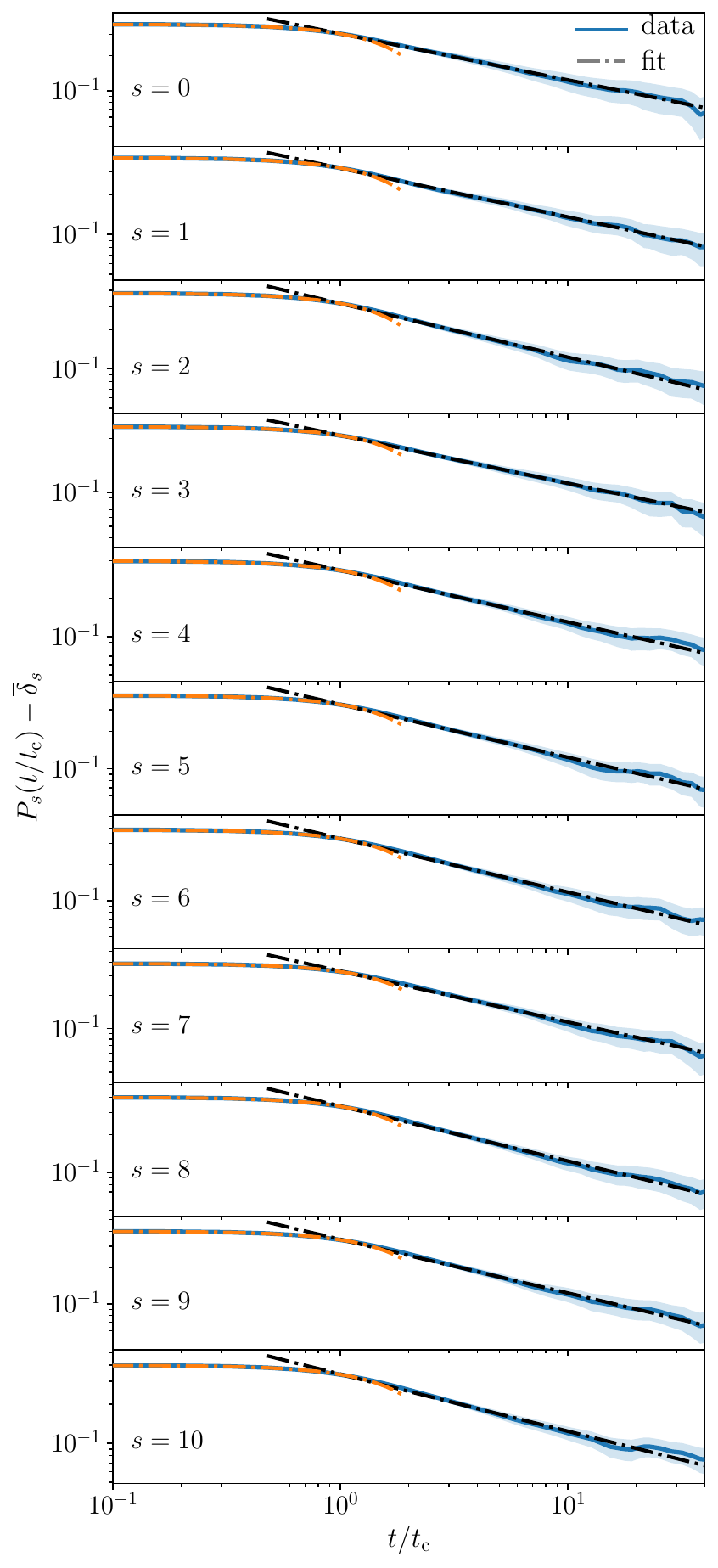}
		\caption{Mean transmitted power $P_s(t/t_\text{c})-\overline{\delta}_s$ for $r_0=15$\,mm.}
		\label{supp:fig15mm}
	\end{figure*}

\begin{figure*}[h!]
		\includegraphics[width=.8\columnwidth]{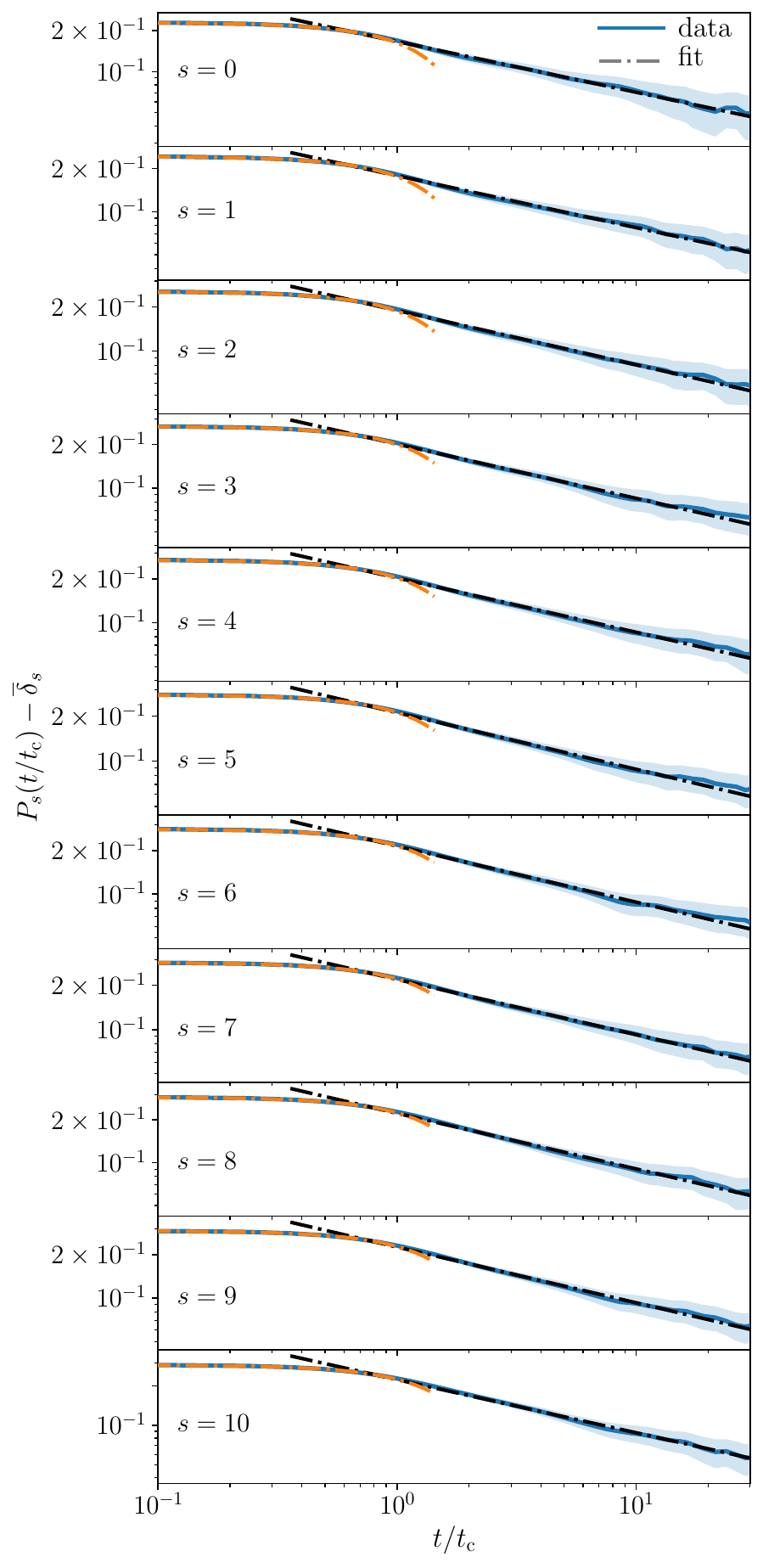}
		\caption{Mean transmitted power $P_s(t/t_\text{c})-\overline{\delta}_s$ for $r_0=20$\,mm.}
		\label{supp:fig20mm}
	\end{figure*}

\begin{figure*}[h!]
		\includegraphics[width=.8\columnwidth]{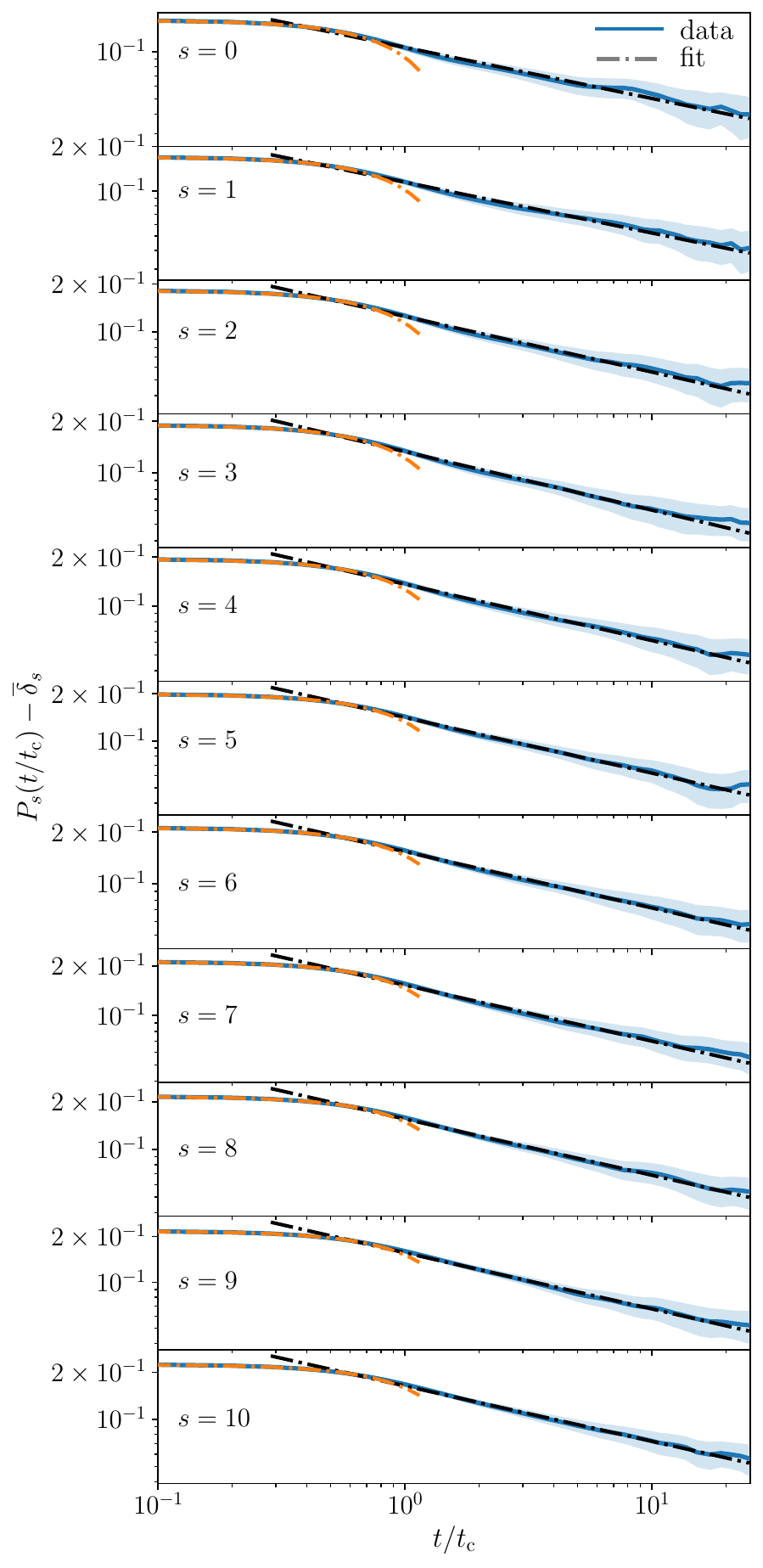}
		\caption{Mean transmitted power $P_s(t/t_\text{c})-\overline{\delta}_s$ for $r_0=25$\,mm.}
		\label{supp:fig25mm}
	\end{figure*}

\begin{figure*}[h!]
		\includegraphics[width=.8\columnwidth]{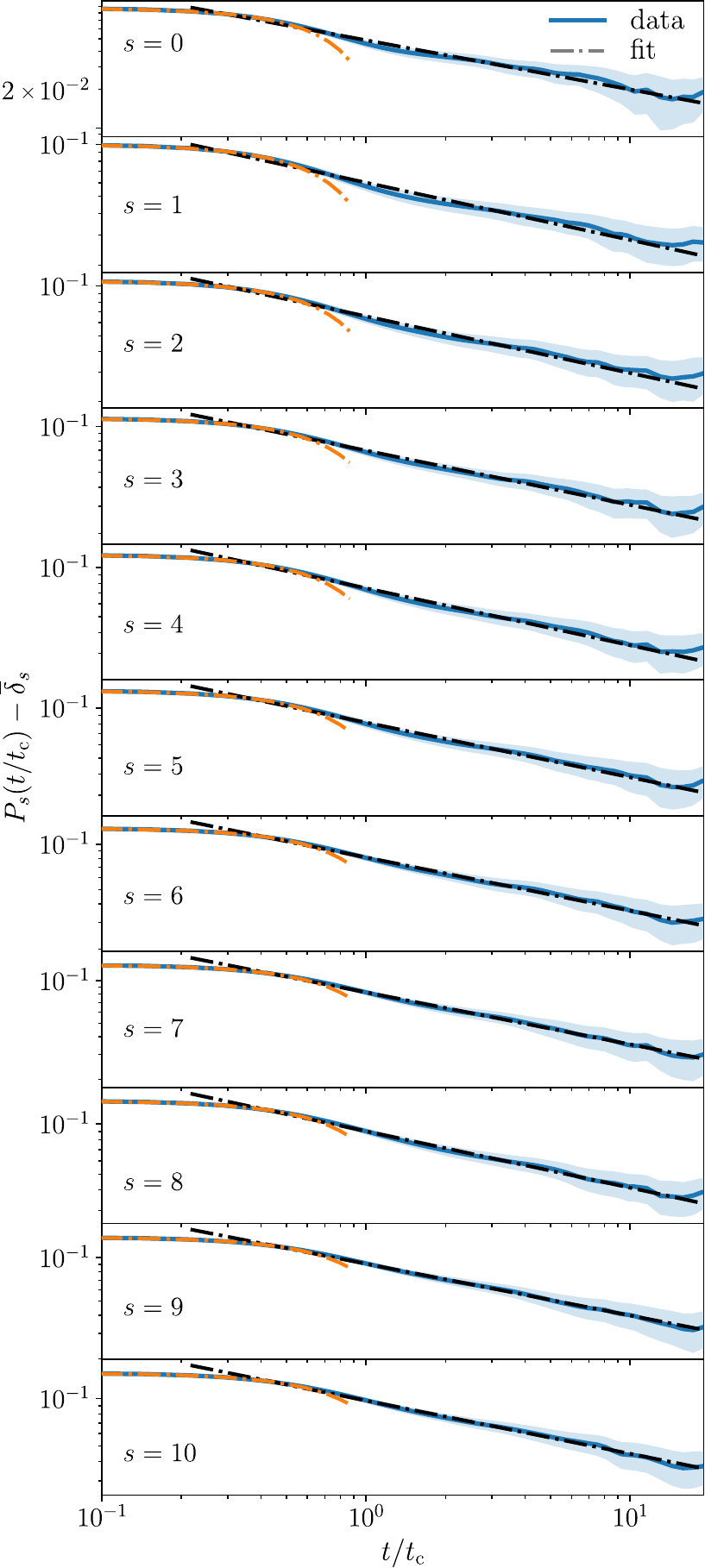}
		\caption{Mean transmitted power $P_s(t/t_\text{c})-\overline{\delta}_s$ for $r_0=33$\,mm.}
		\label{supp:fig33mm}
	\end{figure*}

\begin{figure*}[h!]
		\hspace{-30pt}
		\includegraphics[width=1.2\columnwidth]{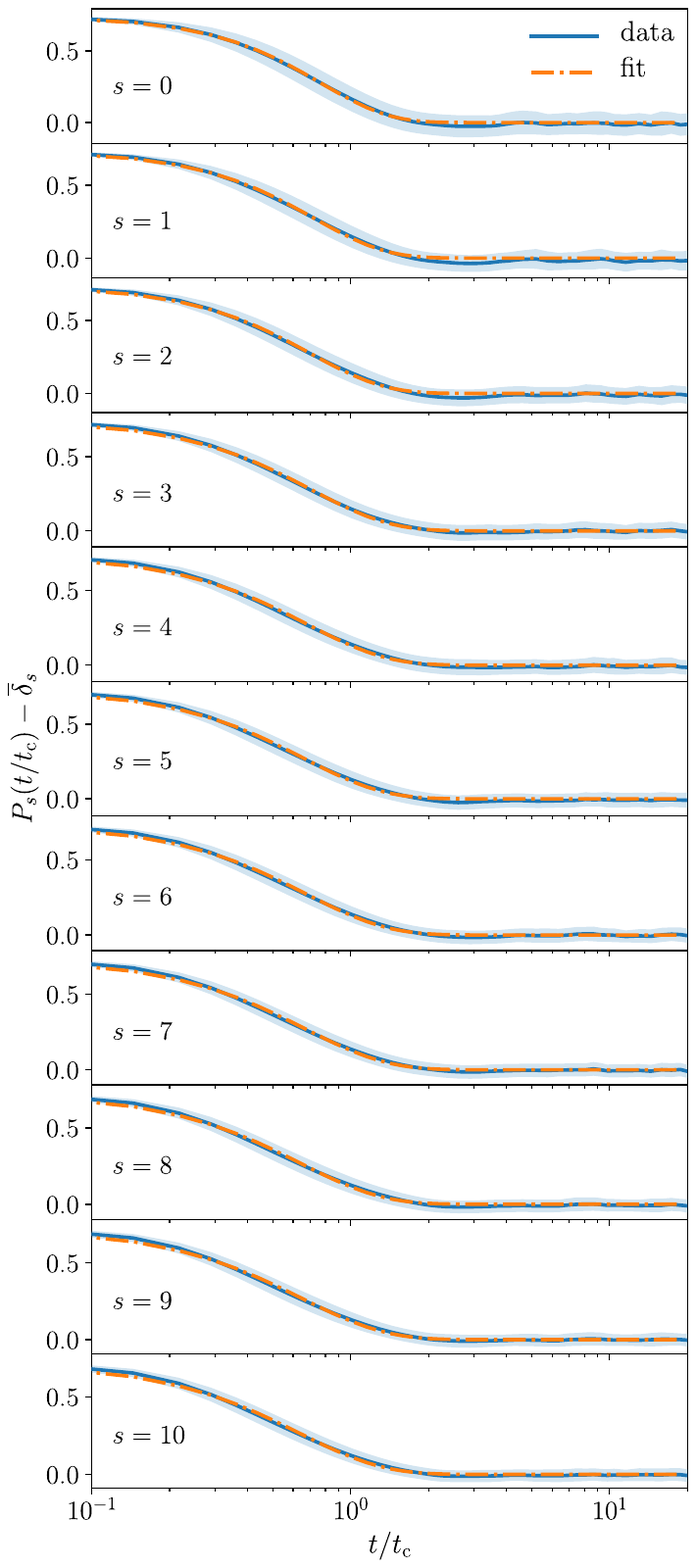}
		\hspace{-60pt}
		\includegraphics[width=1.2\columnwidth]{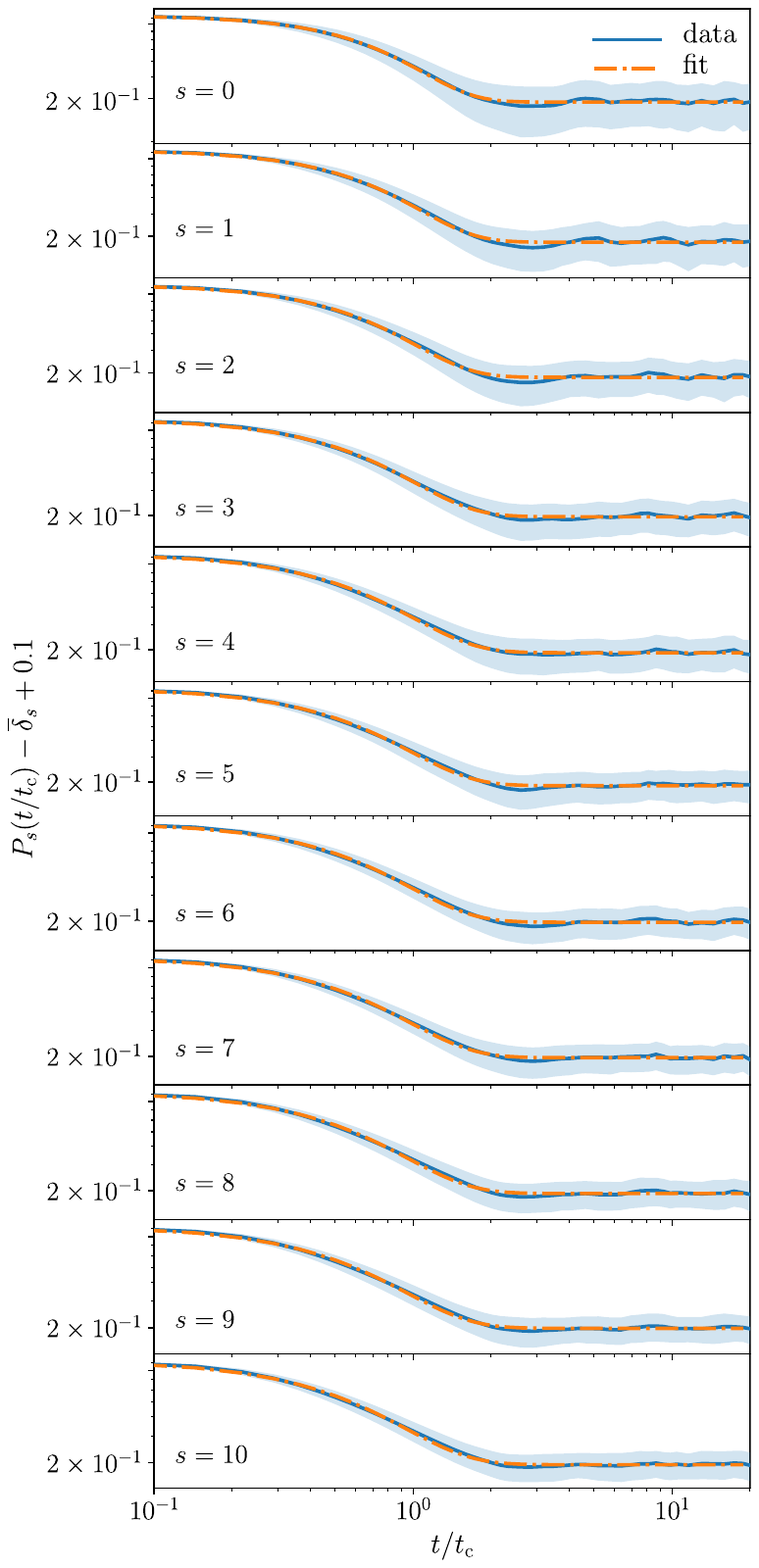}
		\caption{Mean transmitted power $P_s(t/t_\text{c})-\overline{\delta}_s$ for $r=33$\,mm and the Gaussian power spectrum given in Eq.~\eqref{eq:gauss}. Left: Semi-logarithmic plot. Right: Double-Logarithmic plot (an offset of 0.1 was added to avoid zero crossings).}
		\label{supp:fig33mm_ps}
	\end{figure*}

	All simulations of the atmosphere relied on the Kolmogorov power spectrum of turbulence \cite{Kolmogorov41a, Andrews05}
	
	\begin{equation}
		\Phi_n(\kappa) \sim \kappa^{-11/3},
		\label{eq:kol}
	\end{equation}
	where $\kappa$ denotes the transverse spatial frequency of refractive index $n$ fluctuations. In this case, we obtain the initial exponential decay of transmitted power which transits into an algebraic decay for larger times, cf. Figs.~\ref{supp:fig15mm}-\ref{supp:fig33mm} and given in Eq.~\eqref{supp:Pt} [Eq.~(3) in the main manuscript].
	Additionally, we employed a Gaussian (i.e., non-Kolmogorov) power spectrum \cite{Martin88, Ishimaru78}
	\begin{equation}
		\Phi_n(\kappa) \sim r^2 \exp\left(-\kappa^2\,r^2\right),
		\label{eq:gauss}
	\end{equation}
	where we chose a length scale of $r=33\,$mm and seven propagation steps (cf. Table~\ref{supp:table1}) leading to similar turbulence strengths as for the Kolmogorov spectrum with $r_0=33\,$mm. 
	Simulations based on the spectrum given in Eq.~\eqref{eq:gauss} entail exponential decay of the transmitted power with time, cf. Fig.~\ref{supp:fig33mm_ps} and Eq.~\eqref{eq:Ptps}.
	
	To complement Fig.~3 of the main manuscript, Figs.~\ref{supp:fig15mm}-\ref{supp:fig33mm} are plotted on double logarithmic scales in order to highlight the difference of the two decay regimes. The blue solid curve in each subplot shows the numerically obtained mean transmitted power (with its asymptotic offset subtracted), i.e., $P_s(t/t_\text{c})-\overline{\delta}_s$, versus time in units of atmospheric coherence times $t_\text{c}$ corresponding to singular mode $s$.  The offset $\overline{\delta}_s$ is found numerically by ensemble averaging the transmitted power of singular modes associated with a given realization of turbulence through a statistically independent realization with the same Fried parameter $r_0$. The ensemble averaging was performed over at least $300$ disorder realizations and the blue error bands correspond to one standard deviation.
	
	The dot-dashed orange and black curves in Figs.~\ref{supp:fig15mm}-\ref{supp:fig33mm} [atmospheric Kolmogorov spectrum, i.e., Eq.~\eqref{eq:kol}] are obtained by two independent nonlinear fits. Both fitting parameters and their errors are determined with the \texttt{python} library \texttt{scipy.optimize.curve\_fit} and the crossover time $t_\text{x}$ between the two fitting domains is given by the inflection point of $P_s(t/t_\text{c})$. Since our numerical simulation yields discrete data points, $t_\text{x}$ was found by maximization of the difference quotient
	\begin{equation}
		\Big|\frac{P_s(t_{i-1})-P_s(t_i)}{t_{i-1}-t_i}\Big|.
	\end{equation}
	The explicit form of the two fitting functions (composing a single piecewise function) reads
	\begin{equation}
		P_s(t/t_\text{c})= \left\{
		\begin{array}{ll}
			\overline{\tau}_s \exp\left[- ( a_s t/t_\text{c})^{5/3}\right] & \quad t \lesssim t_\text{x}\\
			b_s\,(t/t_\text{c})^{-c_s} + \bar{\delta}_s & \quad t > t_\text{x}. \\
			
		\end{array}
		\right.
		\label{supp:Pt}
	\end{equation}
	Thus, the initial exponential decay [orange dot-dashed curve in Figs.~\ref{supp:fig15mm}-\ref{supp:fig33mm} and first line of Eq.~\eqref{supp:Pt}] for $t \lesssim t_\text{x}$ was fitted with a single parameter $a_s$ (where the index $s$ denotes as usual the singular mode number ordered by their initial transmittance) and $\overline{\tau}_s$ is given by the ensemble average of the modes' corresponding singular values $\tau_s$. Notably, $a_s$ depends neither on the singular mode number $s$, nor on the Fried parameter $r_0$ and, due to time rescaling with $t_\text{c}$, nor on the mean wind velocity $\langle V \rangle$, cf. Tables~\ref{supp:table15mm}-\ref{supp:table33mm}.
	
	The fit of the subsequent algebraic decay [black dot-dashed curve in Figs.~\ref{supp:fig15mm}-\ref{supp:fig33mm} and second line of Eq.~\eqref{supp:Pt}] for $t > t_\text{x}$ relies on two fitting parameters, $b_s$ and $c_s$, while the asymptotic transmittance $\overline{\delta}_s$ was determined independently as described above (see the obtained numerical values in Tables~\ref{supp:table15mm}-\ref{supp:table33mm}). Consistently, if the asymptotic offset was treated as an additional fitting parameter $d_s$, we typically obtained the same value within one standard deviation, cf. the last two columns of Tables~\ref{supp:table15mm}-\ref{supp:table33mm}.
	
	Finally, Fig.~\ref{supp:fig33mm_ps} depicts the singular modes's power transmittance minus $\overline{\delta}_s$ for $r_0=33\,$mm and the Gaussian power spectrum,  Eq.~\eqref{eq:gauss}. As expected, the absence of long-range spatial correlations in this case results in a purely exponential power decay. The left panels of Fig.~\ref{supp:fig33mm_ps} are plotted on semi-logarithmic scales, i.e., only the horizontal time axis is logarithmic, and the right panels of Fig.~\ref{supp:fig33mm_ps} employ double-logarithmic scales (an offset of 0.1 was added in order to avoid zero crossings which are now possible due to the fast convergence of the pure exponential decay).
	The numerical data is shown as blue solid curves with error bands corresponding to one standard deviation of more than $150$ realizations. The exponential fit (dot-dashed orange curve) is of the form
	
	\begin{equation}
		P_s(t/t_\text{c}) = \left(\overline{\tau}_s -\overline{\delta}_s\right) \exp \left[ - \left( a_s t/t_\text{c}\right)^{b_s}\right] + \overline{\delta}_s,
		\label{eq:Ptps}
	\end{equation}
	where $a_s$ and $b_s$ are fitting parameters and $\overline{\tau}_s$ and $\overline{\delta}_s$ were obtained independently as described above (errors are one standard deviation), with numerical values given in Table~\ref{supp:table33mm_ps}. When treating the asymptotic offset as an additional fitting parameter $d_s$, it agrees with $\overline{\delta}_s$ within one standard deviation, cf. last two columns of Table~\ref{supp:table33mm_ps}.
	
	\newpage
	
	\begin{table*}[h!]
		\centering
		\begin{tabular}{r|r|r|r|r|r|r}
			$s$ & $\overline{\tau}_s$ & $a_s$ & $b_s$ & $c_s$ & $d_s$ (fit) & $\overline{\delta}_s$ (uncorr. turbulence)\\\hline\hline
			0 & 0.954 $\pm$ $0.003$ & 0.200 $\pm$ 0.002 & 0.305 $\pm$ 0.001 & 0.391 $\pm$ 0.003 & 0.584 $\pm$ 0.007 & 0.586 $\pm$ 0.010 \\
			1 & 0.951 $\pm$ 0.003 & 0.200 $\pm$ 0.002 & 0.318 $\pm$ 0.001 & 0.370 $\pm$ 0.002 & 0.569 $\pm$ 0.006 & 0.569 $\pm$ 0.010 \\
			2 & 0.949 $\pm$ 0.003 & 0.201 $\pm$ 0.002 & 0.317 $\pm$ 0.002 & 0.411 $\pm$ 0.004 & 0.576 $\pm$ 0.007 & 0.568 $\pm$ 0.009 \\
			3 & 0.948 $\pm$ 0.003 & 0.201 $\pm$ 0.002 & 0.318 $\pm$ 0.001 & 0.423 $\pm$ 0.003 & 0.550 $\pm$ 0.007 & 0.567 $\pm$ 0.008 \\
			4 & 0.946 $\pm$ 0.004 & 0.201 $\pm$ 0.002 & 0.334 $\pm$ 0.002 & 0.408 $\pm$ 0.004 & 0.549 $\pm$ 0.010 & 0.549 $\pm$ 0.009 \\
			5 & 0.944 $\pm$ 0.004 & 0.199 $\pm$ 0.002 & 0.331 $\pm$ 0.002 & 0.428 $\pm$ 0.005 & 0.545 $\pm$ 0.010 & 0.553 $\pm$ 0.008 \\
			6 & 0.943 $\pm$ 0.004 & 0.199 $\pm$ 0.002 & 0.329 $\pm$ 0.002 & 0.446 $\pm$ 0.005 & 0.545 $\pm$ 0.010 & 0.554 $\pm$ 0.008 \\
			7 & 0.942 $\pm$ 0.004 & 0.198 $\pm$ 0.002 & 0.331 $\pm$ 0.002 & 0.462 $\pm$ 0.006 & 0.525 $\pm$ 0.011 & 0.553 $\pm$ 0.008 \\
			8 & 0.940 $\pm$ 0.004 & 0.200 $\pm$ 0.002 & 0.338 $\pm$ 0.002 & 0.438 $\pm$ 0.005 & 0.534 $\pm$ 0.010 & 0.542 $\pm$ 0.008 \\
			9 & 0.939 $\pm$ 0.005 & 0.198 $\pm$ 0.002 & 0.342 $\pm$ 0.002 & 0.444 $\pm$ 0.005 & 0.525 $\pm$ 0.011 & 0.539 $\pm$ 0.008 \\
			10 & 0.937 $\pm$ 0.005 & 0.197 $\pm$ 0.002 & 0.340 $\pm$ 0.003 & 0.439 $\pm$ 0.006 & 0.535 $\pm$ 0.011 & 0.538 $\pm$ 0.008 \\
		\end{tabular}
		\caption{Parameters of Eq.~\eqref{supp:Pt} for $r_0=15$\,mm.}
		\label{supp:table15mm}
	\end{table*}

	\newpage
	
	\begin{table*}[h!]
		\centering
		\begin{tabular}{r|r|r|r|r|r|r}
			$s$ & $\overline{\tau}_s$ & $a_s$ & $b_s$ & $c_s$ & $d_s$ (fit) & $\overline{\delta}_s$ (uncorr. turbulence)\\\hline\hline
			0 & 0.972 $\pm$ 0.001 & 0.205 $\pm$ 0.002 & 0.166 $\pm$ 0.001 & 0.372 $\pm$ 0.003 & 0.751 $\pm$ 0.003 & 0.743 $\pm$ 0.008 \\
			1 & 0.971 $\pm$ 0.002 & 0.207 $\pm$ 0.002 & 0.178 $\pm$ 0.001 & 0.363 $\pm$ 0.003 & 0.744 $\pm$ 0.003 & 0.727 $\pm$ 0.008 \\
			2 & 0.969 $\pm$ 0.002 & 0.206 $\pm$ 0.002 & 0.189 $\pm$ 0.001 & 0.370 $\pm$ 0.003 & 0.730 $\pm$ 0.003 & 0.715 $\pm$ 0.007 \\
			3 & 0.968 $\pm$ 0.002 & 0.206 $\pm$ 0.002 & 0.201 $\pm$ 0.001 & 0.374 $\pm$ 0.003 & 0.712 $\pm$ 0.005 & 0.701 $\pm$ 0.007 \\
			4 & 0.967 $\pm$ 0.002 & 0.206 $\pm$ 0.002 & 0.203 $\pm$ 0.001 & 0.373 $\pm$ 0.003 & 0.715 $\pm$ 0.004 & 0.698 $\pm$ 0.007 \\
			5 & 0.966 $\pm$ 0.002 & 0.205 $\pm$ 0.002 & 0.211 $\pm$ 0.001 & 0.377 $\pm$ 0.004 & 0.703 $\pm$ 0.006 & 0.689 $\pm$ 0.007 \\
			6 & 0.966 $\pm$ 0.002 & 0.204 $\pm$ 0.002 & 0.214 $\pm$ 0.001 & 0.386 $\pm$ 0.004 & 0.692 $\pm$ 0.006 & 0.685 $\pm$ 0.007 \\
			7 & 0.965 $\pm$ 0.002 & 0.202 $\pm$ 0.002 & 0.219 $\pm$ 0.001 & 0.376 $\pm$ 0.003 & 0.683 $\pm$ 0.006 & 0.679 $\pm$ 0.007 \\
			8 & 0.964 $\pm$ 0.002 & 0.203 $\pm$ 0.002 & 0.223 $\pm$ 0.001 & 0.391 $\pm$ 0.004 & 0.678 $\pm$ 0.007 & 0.674 $\pm$ 0.007 \\
			9 & 0.963 $\pm$ 0.003 & 0.202 $\pm$ 0.002 & 0.226 $\pm$ 0.001 & 0.386 $\pm$ 0.004 & 0.675 $\pm$ 0.007 & 0.671 $\pm$ 0.007 \\
			10 & 0.962 $\pm$ 0.003 & 0.202 $\pm$ 0.002 & 0.223 $\pm$ 0.001 & 0.405 $\pm$ 0.005 & 0.656 $\pm$ 0.009 & 0.673 $\pm$ 0.007 \\
		\end{tabular}
		\caption{Parameters of Eq.~\eqref{supp:Pt} for $r_0=20$\,mm.}
		\label{supp:table20mm}
	\end{table*}

	\newpage 
	
	\begin{table*}[h!]
		\centering
		\begin{tabular}{r|r|r|r|r|r|r}
			$s$ & $\overline{\tau}_s$ & $a_s$ & $b_s$ & $c_s$ & $d_s$ (fit) & $\overline{\delta}_s$ (uncorr. turbulence)\\\hline\hline
			0 & 0.981 $\pm$ 0.001 & 0.205 $\pm$ 0.002 & 0.107 $\pm$ 0.001 & 0.326 $\pm$ 0.006 & 0.849 $\pm$ 0.002 & 0.822 $\pm$ 0.006 \\
			1 & 0.980 $\pm$ 0.001 & 0.206 $\pm$ 0.002 & 0.115 $\pm$ 0.001 & 0.341 $\pm$ 0.005 & 0.834 $\pm$ 0.002 & 0.810 $\pm$ 0.006 \\
			2 & 0.979 $\pm$ 0.001 & 0.207 $\pm$ 0.002 & 0.125 $\pm$ 0.001 & 0.348 $\pm$ 0.004 & 0.818 $\pm$ 0.002 & 0.797 $\pm$ 0.006 \\
			3 & 0.978 $\pm$ 0.001 & 0.207 $\pm$ 0.002 & 0.133 $\pm$ 0.001 & 0.341 $\pm$ 0.003 & 0.807 $\pm$ 0.003 & 0.788 $\pm$ 0.006 \\
			4 & 0.978 $\pm$ 0.001 & 0.207 $\pm$ 0.002 & 0.136 $\pm$ 0.001 & 0.346 $\pm$ 0.003 & 0.799 $\pm$ 0.003 & 0.783 $\pm$ 0.006 \\
			5 & 0.977 $\pm$ 0.001 & 0.205 $\pm$ 0.002 & 0.141 $\pm$ 0.001 & 0.356 $\pm$ 0.003 & 0.781 $\pm$ 0.004 & 0.778 $\pm$ 0.006 \\
			6 & 0.976 $\pm$ 0.001 & 0.204 $\pm$ 0.002 & 0.154 $\pm$ 0.001 & 0.330 $\pm$ 0.003 & 0.777 $\pm$ 0.004 & 0.764 $\pm$ 0.006 \\
			7 & 0.976 $\pm$ 0.001 & 0.204 $\pm$ 0.002 & 0.153 $\pm$ 0.001 & 0.340 $\pm$ 0.004 & 0.782 $\pm$ 0.004 & 0.763 $\pm$ 0.006 \\
			8 & 0.975 $\pm$ 0.001 & 0.205 $\pm$ 0.002 & 0.156 $\pm$ 0.001 & 0.355 $\pm$ 0.003 & 0.764 $\pm$ 0.005 & 0.759 $\pm$ 0.006 \\
			9 & 0.975 $\pm$ 0.001 & 0.203 $\pm$ 0.002 & 0.157 $\pm$ 0.001 & 0.368 $\pm$ 0.003 & 0.756 $\pm$ 0.006 & 0.758 $\pm$ 0.006 \\
			10 & 0.974 $\pm$ 0.002 & 0.203 $\pm$ 0.002 & 0.163 $\pm$ 0.001 & 0.353 $\pm$ 0.003 & 0.751 $\pm$ 0.006 & 0.750 $\pm$ 0.006 \\
		\end{tabular}
		\caption{Parameters of Eq.~\eqref{supp:Pt} for $r_0=25$\,mm.}
		\label{supp:table25mm}
	\end{table*}

	\newpage 
	
	\begin{table*}[h!]
		\centering
		\begin{tabular}{r|r|r|r|r|r|r}
			$s$ & $\overline{\tau}_s$ & $a_s$ & $b_s$ & $c_s$ & $d_s$ (fit) & $\overline{\delta}_s$ (uncorr. turbulence)\\\hline\hline
			0 & 0.988 $\pm$ 0.001 & 0.204 $\pm$ 0.002 & 0.049 $\pm$ 0.001 & 0.387 $\pm$ 0.008 & 0.912 $\pm$ 0.002 & 0.901 $\pm$ 0.003 \\
			1 & 0.988 $\pm$ 0.001 & 0.207 $\pm$ 0.002 & 0.060 $\pm$ 0.001 & 0.332 $\pm$ 0.007 & 0.906 $\pm$ 0.002 & 0.887 $\pm$ 0.004 \\
			2 & 0.987 $\pm$ 0.001 & 0.206 $\pm$ 0.002 & 0.066 $\pm$ 0.001 & 0.344 $\pm$ 0.006 & 0.896 $\pm$ 0.002 & 0.879 $\pm$ 0.004 \\
			3 & 0.987 $\pm$ 0.001 & 0.209 $\pm$ 0.002 & 0.070 $\pm$ 0.001 & 0.357 $\pm$ 0.005 & 0.886 $\pm$ 0.002 & 0.873 $\pm$ 0.004 \\
			4 & 0.986 $\pm$ 0.001 & 0.209 $\pm$ 0.002 & 0.075 $\pm$ 0.001 & 0.349 $\pm$ 0.005 & 0.881 $\pm$ 0.002 & 0.866 $\pm$ 0.004 \\
			5 & 0.986 $\pm$ 0.001 & 0.207 $\pm$ 0.002 & 0.082 $\pm$ 0.001 & 0.328 $\pm$ 0.004 & 0.874 $\pm$ 0.003 & 0.859 $\pm$ 0.005 \\
			6 & 0.985 $\pm$ 0.001 & 0.206 $\pm$ 0.002 & 0.082 $\pm$ 0.001 & 0.356 $\pm$ 0.004 & 0.862 $\pm$ 0.003 & 0.857 $\pm$ 0.004 \\
			7 & 0.985 $\pm$ 0.001 & 0.205 $\pm$ 0.002 & 0.083 $\pm$ 0.001 & 0.370 $\pm$ 0.003 & 0.854 $\pm$ 0.003 & 0.855 $\pm$ 0.004 \\
			8 & 0.985 $\pm$ 0.001 & 0.206 $\pm$ 0.002 & 0.090 $\pm$ 0.001 & 0.341 $\pm$ 0.003 & 0.854 $\pm$ 0.003 & 0.847 $\pm$ 0.005 \\
			9 & 0.984 $\pm$ 0.001 & 0.204 $\pm$ 0.002 & 0.091 $\pm$ 0.001 & 0.357 $\pm$ 0.003 & 0.839 $\pm$ 0.004 & 0.846 $\pm$ 0.004 \\
			10 & 0.984 $\pm$ 0.001 & 0.204 $\pm$ 0.002 & 0.097 $\pm$ 0.001 & 0.338 $\pm$ 0.003 & 0.835 $\pm$ 0.005 & 0.838 $\pm$ 0.004 \\
		\end{tabular}
		\caption{Parameters of Eq.~\eqref{supp:Pt} for $r_0=33$\,mm.}
		\label{supp:table33mm}
	\end{table*}

	\begin{table*}[h!]
		\centering
		
		\begin{tabular}{r|r|r|r|r|r}
			$s$ & $\overline{\tau}_s$ & $a_s$ & $b_s$ & $d_s$ (fit) & $\overline{\delta}_s$ (uncorr. turbulence) \\\hline\hline
			0 & 0.995 $\pm$ 0.005 & 1.295 $\pm$ 0.012 & 1.665 $\pm$ 0.037 & 0.247 $\pm$ 0.001 & 0.254 $\pm$ 0.013 \\
			1 & 0.991 $\pm$ 0.008 & 1.377 $\pm$ 0.019 & 1.593 $\pm$ 0.053 & 0.240 $\pm$ 0.002 & 0.255 $\pm$ 0.010 \\
			2 & 0.987 $\pm$ 0.011 & 1.424 $\pm$ 0.016 & 1.541 $\pm$ 0.041 & 0.237 $\pm$ 0.002 & 0.246 $\pm$ 0.009 \\
			3 & 0.983 $\pm$ 0.013 & 1.408 $\pm$ 0.009 & 1.450 $\pm$ 0.022 & 0.237 $\pm$ 0.002 & 0.235 $\pm$ 0.008 \\
			4 & 0.979 $\pm$ 0.016 & 1.464 $\pm$ 0.013 & 1.424 $\pm$ 0.028 & 0.236 $\pm$ 0.002 & 0.240 $\pm$ 0.009 \\
			5 & 0.974 $\pm$ 0.018 & 1.509 $\pm$ 0.016 & 1.435 $\pm$ 0.033 & 0.234 $\pm$ 0.002 & 0.240 $\pm$ 0.008 \\
			6 & 0.969 $\pm$ 0.022 & 1.477 $\pm$ 0.011 & 1.398 $\pm$ 0.023 & 0.234 $\pm$ 0.002 & 0.232 $\pm$ 0.007 \\
			7 & 0.965 $\pm$ 0.024 & 1.494 $\pm$ 0.011 & 1.384 $\pm$ 0.022 & 0.231 $\pm$ 0.002 & 0.229 $\pm$ 0.007 \\
			8 & 0.960 $\pm$ 0.027 & 1.555 $\pm$ 0.015 & 1.388 $\pm$ 0.029 & 0.232 $\pm$ 0.002 & 0.235 $\pm$ 0.007 \\
			9 & 0.954 $\pm$ 0.029 & 1.534 $\pm$ 0.011 & 1.356 $\pm$ 0.021 & 0.230 $\pm$ 0.002 & 0.227 $\pm$ 0.006 \\
			10 & 0.949 $\pm$ 0.032 & 1.569 $\pm$ 0.013 & 1.361 $\pm$ 0.024 & 0.230 $\pm$ 0.002 & 0.230 $\pm$ 0.006 \\
		\end{tabular}
		\caption{Parameters of Eq.~\eqref{eq:Ptps} (Gaussian power spectrum) for $r=33$\,mm.}
		\label{supp:table33mm_ps}
	\end{table*}
	
\end{document}